\documentclass[%
 aip,
 amsmath,amssymb,
 reprint,%
]{revtex4-1}

\usepackage{graphicx}
\usepackage{dcolumn}
\usepackage{bm}

\usepackage[utf8]{inputenc}
\usepackage[T1]{fontenc}
\usepackage{mathptmx}
\usepackage{etoolbox}

\makeatletter
\def\@email#1#2{%
 \endgroup
 \patchcmd{\titleblock@produce}
  {\frontmatter@RRAPformat}
  {\frontmatter@RRAPformat{\produce@RRAP{*#1\href{mailto:#2}{#2}}}\frontmatter@RRAPformat}
  {}{}
}%
\makeatother
\begin{document}

\preprint{AIP/123-QED}

\title[Title]{On the flow characteristics in the shock formation region due to the diaphragm opening process in a shock tube}

\author{Touqeer Anwar Kashif}
\affiliation{Mechanical Engineering, Physical Science and Engineering Division, King Abdullah University of Science and Technology (KAUST), Thuwal, 23955-6900, Kingdom of Saudi Arabia}
\author{Janardhanraj Subburaj}
\affiliation{Mechanical Engineering, Physical Science and Engineering Division, King Abdullah University of Science and Technology (KAUST), Thuwal, 23955-6900, Kingdom of Saudi Arabia}
  \email{janardhanraj.subburaj@kaust.edu.sa}
\author{A. Farooq}
\affiliation{Mechanical Engineering, Physical Science and Engineering Division, King Abdullah University of Science and Technology (KAUST), Thuwal, 23955-6900, Kingdom of Saudi Arabia}

\date{\today}

\begin{abstract}
The shock formation process in shock tubes has been extensively studied; however, significant gaps remain in understanding the effects of the diaphragm rupture process on the resulting flow non-uniformities. Existing models predicting the shock attenuation and propagation dynamics overlook critical diaphragm mechanics and their impact on shock behavior. Addressing this gap is vital for improving predictive capabilities and optimizing shock tube designs for applications in combustion kinetics, aerodynamics, and high-speed diagnostics. This study investigates the shock wave formation through combined experimental and numerical approaches over a range of driver-to-driven pressure ratios (Driver pressure: 9.4 - 25.5 bar of helium; Driven pressure: 100 Torr (133.322 mbar) of argon). High-speed imaging captures the diaphragm opening dynamics, while pressure and shock velocity measurements along the entire driven section of the shock tube provide key validation data for computational fluid dynamic simulations. Two-dimensional numerical simulations incorporate experimentally measured diaphragm opening profiles, offering detailed insights into flow features and thermodynamic gradients behind the moving shock front. Key parameters, including deceleration and acceleration phases within the shock formation region, shock formation distances, and times, have been quantified. A novel theoretical framework is introduced to correlate these parameters, enabling accurate predictions of shock Mach number evolution under varying conditions. This unified methodology bridges theoretical and experimental gaps, providing a robust foundation for advancing shock tube research and design. 
\end{abstract}

\maketitle

\section{\label{sec:intro}INTRODUCTION}
Shock wave research has been pivotal over many decades in advancing diverse fields, including aerodynamics, hypersonics, chemical kinetics, and combustion.\cite{Gu_2020,Bhaskaran_2002,Reynier_2016} Its applications have increasingly extended to interdisciplinary domains such as materials science,\cite{ana@2022,Maity_2019,Bisht_2019} medicine,\cite{Subburaj_2017} bioengineering,\cite{Datey_2017,Gnanadhas_2015} geophysics,\cite{Takayama_2004} and industrial processes.\cite{hariharan@2011} Shock tubes are among the most commonly used experimental facilities for studying shock wave phenomena and gaining insights into modern gas dynamics. Consequently, there has been longstanding interest in understanding the formation of shock waves in shock tubes and the complex flow dynamics that ensue. A standard shock tube consists of two cylindrical sections, the driver and the driven sections, separated by a diaphragm. The driver section is typically filled with a light gas, such as hydrogen or helium, while the driven section contains the test gas, selected based on the specific application. For example, atmospheric re-entry studies use the atmospheric composition of the target planet as the test gas, while combustion studies often involve mixtures of fuel, oxidizer, and diluents. When the diaphragm ruptures due to the pressure differential between the two sections, the driver gas rapidly expands into the driven section, generating a shock wave. The thermodynamic state of the gas on both sides of the shock wave is governed by the Rankine-Hugoniot relations.\cite{Needham_2018} Additionally, the ideal shock tube theory (Eq. \ref{eq:ideal_relation}) establishes a relationship between the initial thermodynamic conditions and the shock Mach number (the ratio of the shock speed to the speed of sound in the medium), under assumptions of one-dimensional, inviscid, adiabatic flow, instantaneous diaphragm rupture, and shock wave formation at the diaphragm location. 

\begin{equation}
\resizebox{.9\hsize}{!}{$P_{41} = \frac{2\gamma_1 M_{s,ideal}^2 - (\gamma_1 - 1)}{\gamma_1 + 1} \left( 1 - \frac{\gamma_4 - 1}{\gamma_1 + 1}\frac{a_1}{a_4} \left( M_{s,ideal} - \frac{1}{M_{s,ideal}} \right) \right)^{\frac{-2\gamma_4}{\gamma_4-1}}$}
\label{eq:ideal_relation}
\end{equation}

where $P_{41}$ is the pressure ratio of driver to driven gas, likewise $a$ and $\gamma$ denote the speed of sound and specific heat of the gas, with subscript 4 and 1 indicating driver gas and driven gases respectively. $M_{s,ideal}$ is the Mach number of the shock wave.
Although developed decades ago, the ideal shock tube relation is frequently referenced to estimate shock strength under given initial conditions. However, the predictions often deviate significantly from experimental measurements due to the idealized assumptions that do not accurately represent shock tube operation. Two primary factors contributing to these deviations are losses during the diaphragm opening process and shock attenuation caused by boundary layer growth.\cite{ikui1969investigations,ikui1969investigations1} Therefore, there are several studies that focus on the independent influence of these factors on the post-shock conditions. 

Early investigations into the diaphragm opening process examined both high-strength materials, such as aluminum, stainless steel, and copper, and lower-strength materials, including Mylar and polycarbonate. Research has shown that diaphragm rupture is a gradual process that significantly influences the initial shock’s intensity and velocity. White\cite{white1958influence} demonstrated that the shock wave continues to accelerate until the diaphragm is fully open, reaching its peak velocity at that point. This study used a photomultiplier tube to observe a light source placed at the driver end, recording opening times of approximately 600 microseconds and noting that it took about 200 microseconds for the diaphragm to open from 0 to 10\% of its final value. Similarly, Campbell et al.\cite{campbell1965bursting} utilized both cameras and photoelectric methods to determine opening times for materials like aluminum and copper, while Simpson et al.\cite{simpson1967effect} explored the use of soft nickel to achieve faster opening times. In another study, Rothkopf et al.\cite{rothkopf1974diaphragm} measured diaphragm opening times under pressures exceeding 38 bar using photographic and photomultiplier setups. Collectively, these studies suggest that the slow diaphragm opening prolongs the shock wave's acceleration phase, leading to a higher peak shock velocity. This, in turn, results in stronger shock attenuation due to the accelerated growth of the boundary layer behind the faster shock wave  as reported by White\cite{white1958influence} and Rothkopf et al.\cite{rothkopf1976shock}. These studies also revealed that at high driver-to-driven pressure ratios ($P_{41}$ $\geq$ 500) the peak shock velocities often exceed those predicted by the ideal shock tube theory. Subsequently, the shock wave undergoes linear deceleration due to the growing boundary layer behind it, which acts as a mass sink and produces expansion waves that decelerate the shock wave. Boundary layer growth and its effects on shock attenuation have also been explored extensively. Theoretical works by Mirels\cite{mirels1963test,mirels1957attenuation,mirels1961laminar,mirels1964shock} provide methods to estimate boundary layer thickness in both laminar and turbulent regimes and its impact on test time and shock attenuation. Updated expressions for turbulent boundary layer growth were proposed by Petersen\cite{petersen2003improved}. However, few studies have addressed the coupled effects of diaphragm dynamics and boundary layer growth.

Early numerical studies that incorporated diaphragm opening dynamics in the simulations assumed inviscid boundary conditions. The study by  Petrie-Repar et al.\cite{petrie1998computational} utilized an in-house code to solve the inviscid Euler equations in a 2D-axisymmetric configuration. The diaphragm opening was modeled as a dilating iris with a consistent opening time of 200 microseconds across all investigated pressure ratios. Gaetani et al.\cite{gaetani2008shock} conducted simulations of flow past partially opened diaphragms using an axisymmetric inviscid solver, focusing on relatively low pressure ratios and highlighting the influence of a partially opened diaphragm. However, these studies often failed to fully capture the flow dynamics during shock formation and attenuation that are influenced significantly by the diaphragm opening process. A few recent studies have integrated the diaphragm opening process with viscous boundary conditions. Satchell et al.\cite{satchell2021numerical,satchell2022flow} developed a viscous shock tube solver with shock tracking capabilities, validated against experimental data for a range of Mach numbers, and incorporated a gradual diaphragm opening modeled similar to Petrie-Repar et al.\cite{petrie1998computational} approach (i.e., a dilating iris). Andreotti et al.\cite{andreotti2015performance} conducted one of the most comprehensive shock-tunnel simulations, which included modeling of tube vibratory responses and fluid-structure interactions involving the diaphragm's metal, showing general agreement between 3-D and 1-D inviscid simulations. Moreover, Currao et al.\cite{currao2024diaphragm} performed 2-D and 3-D simulations of double-diaphragm operated shock tubes in ANSYS Fluent, incorporating a gradual dilating iris-like diaphragm opening. Their simulations captured the shock attenuation well when the diaphragm opening process was modeled, although significant oscillations in shock velocity were observed. Lamnaouer et al.\cite{lamnaouer2010time} conducted axisymmetric simulations that matched experimental data in the reflected shock region but did not address the extent of shock attenuation or diaphragm dynamics. Liu et al.\cite{liu2024general} proposed a general theory of diaphragm rupture in shock tunnels that was validated against experimental data from literature. They presented predictions for two types of diaphragm opening patterns, namely ideal and non-ideal diaphragm rupture.

Building on decades of shock tube research, the present study addresses critical gaps in the understanding of shock formation dynamics, particularly the effects of diaphragm rupture on the flow non-uniformities behind the shock front. High-fidelity numerical simulations, validated against experimental measurements, capture the transient shock wave behavior and reveal underlying flow features that are challenging to measure directly. A detailed analysis of diaphragm opening dynamics and its impact on shock velocity profiles, formation distances, and thermodynamic properties forms the core of this work. Moreover, this study introduces a predictive framework for shock Mach number evolution, enabling improved shock wave characterization and experimental design across a broad range of operating conditions. The manuscript is structured as follows: Section II outlines the experimental setup and measurement techniques; Section III describes the numerical methods; Section IV delves into the flow features near the diaphragm and presents a representative shock trajectory; Section V develops correlations for shock parameters; Conclusions and future research directions are discussed in Section VI.

\section {Experimental methods and measurements}

\begin{figure*}
    \centering
    \includegraphics[width=0.95\textwidth]{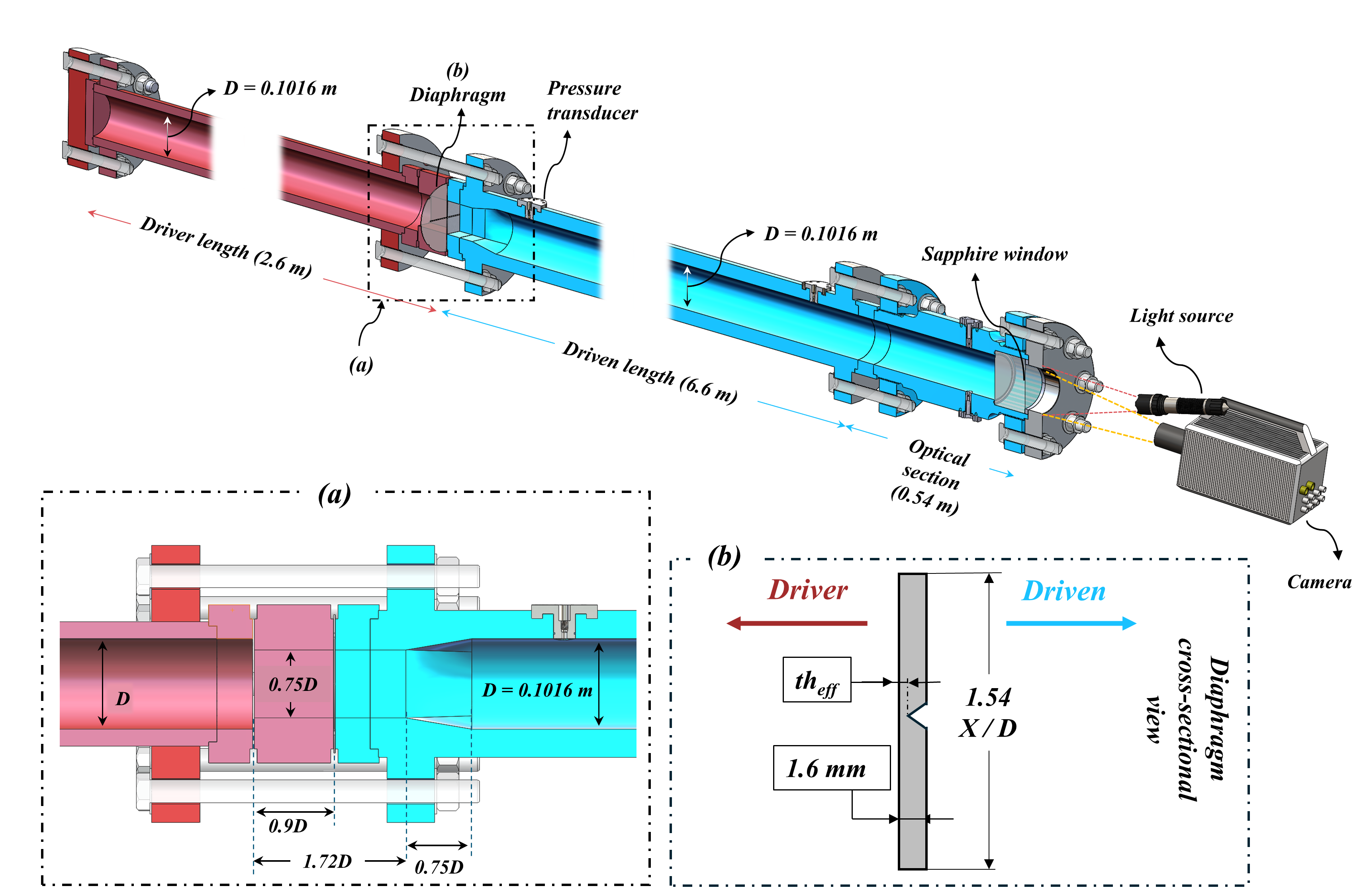}
    \caption{Schematic of the high-pressure shock tube facility at KAUST. A light source placed at driven end illuminates the diaphragm and a high-speed camera placed at the endwall captures the images of the diaphragm. (a) A zoomed cross-sectional view of the area change near the diaphragm station. (b) Cross-sectional cut at the center-line of 1.6 mm thick aluminum diaphragm highlighting the effective thickness of the diaphragm after an 'X' shaped indentation.}
    \label{fig:Schematic_shock_tube}
\end{figure*}

\subsection {Shock tube and imaging setup}
Shock tube experiments were conducted in the High-Pressure Shock Tube (HPST) facility at King Abdullah University of Science and Technology (KAUST). The HPST is constructed from high-quality stainless steel and is designed to withstand static pressures of up to 300 bar. For the present work, the driven section is 7.14 m long and the driver section is 2.6 m with internal diameters of 101.6 mm. The final 0.54 m of the driven section is an optical section with a sapphire end wall for visualizing the diaphragm opening process. A schematic cross-section of the HPST is shown in Fig. \ref{fig:Schematic_shock_tube}. The HPST can operate in both single- and double-diaphragm modes but for the present study only single-diaphragm mode was utilized. The cross-sectional view of the square diaphragm station is shown in Fig. \ref{fig:Schematic_shock_tube}a. An additional square-to-round transition section in the driven section ensures that the flow remains streamlined from the square to circular cross-section. Helium was used as the driver gas and argon as the driven gas, both supplied by AirLiquide Inc. with 99.999 $\%$ purity. The driver pressures were varied between 9.4 and 25.5 bar of Helium, while the driven pressure was kept constant at 100 Torr (133.322 mbar), as outlined in Table \ref{tab:Experimental conditions}. Cases S1 - S5, had 1 bar of air in the driver section before pressurization with Helium, while for S6 and S7 the driver was vacuumed before pressurization. The diaphragms are made of aluminum sheets with a thickness of 1.6 mm and pre-scored in an `X' pattern to ensure symmetrical and debris-free opening were used for the experiments. In Fig. \ref{fig:Schematic_shock_tube}b, a cross-sectional view of the diaphragm is shown and the effective thickness $(t_e)$ is tabulated in Table \ref{tab:Experimental conditions}. 

The trajectory of the shock wave is tracked using twelve pressure transducers (Model: 112B05, PCB Piezotronics, USA), positioned at the axial locations listed in Table \ref{tab:PCB_location}. A layer of silicone RTV was applied to the exposed surface of the transducers for thermal protection. These pressure sensors, with a response time of 1 $\mu$s, were integrated into a data acquisition system (National Instruments, USA). The uncertainty in shock velocity measurements is estimated to be less than 1 \%. Further details on the HPST facility and uncertainty calculations can be found elsewhere. \citep{alabbad2023high,kashif2023effect,figueroa2023dual} The measured shock velocity serves as the basis for calculating key thermodynamic properties, such as temperature and density, behind both the incident and reflected shock waves, using ideal shock relations. Peak Mach number ($M_{s,peak,exp}$) and Mach number at 6.6 m from the diaphragm ($M_{s,endwall,exp}$) acquired experimentally are also tabulated in Table \ref{tab:Experimental conditions}. Likewise pressure behind the incident shock wave at the peak Mach location and at 6.6 m downstream to diaphragm, acquired from ideal shock relations, are also given in Table \ref{tab:Experimental conditions}. A high-speed camera (Model: SA-X2, Photron Limited, Japan) recorded images at 150,000 frames per second, achieving a temporal resolution of 6.6 $\mu$s. The pixel resolution was approximately 1 mm in both the $x$ and $y$ directions. Additionally, the camera was equipped with a 70-200 mm focal lens to view the diaphragm, placed over 8 m away from the camera.  Post-processing involved applying a Canny edge detection filter in MATLAB to identify the diaphragm edges, followed by image binarization to calculate the opened area based on pixel measurements. At very low aperture opened the edge detection couldn't work well, so manual edge tracking was performed in MATLAB. A detailed discussion of the observed diaphragm rupture dynamics is provided in the forthcoming section. 

\begin{table*}
    \def~{\hphantom{0}}
    \begin{center}
    \begin{tabular}{ccccccccc} 
         Notation&  \multicolumn{2}{c}{Driver Pressure, bar}&  $P_{41}$ & $M_{s,peak,exp}$ & $M_{s,endwall,exp}$& $P_{2,peak}$, bar& $P_{2,endwall}$, bar \\ 
 & $(He)$& $(N_2)$ \\[3pt]
 \hline
         S1&  8.4&  1&    70&  3.04&    2.91&  1.51& 1.38 \\ 
         S2&  9.03&  1&    75&  3.10&   2.98&  1.57& 1.45 \\ 
         S3&  11.65&  1&    95&  3.27&   3.18&  1.75& 1.65 \\ 
         S4&  14.75&  1&    118&  3.55&   3.39&  2.07& 1.88 \\ 
         S5&  16.82&  1&    133&  3.64&   3.49&  2.18& 2 \\ 
         S6&  20.13&  0&    151&  4&   3.82&  2.64& 2.4 \\ 
         S7&  25.5&  0&    191&  4.21&   4.02&  2.96& 2.6 \\ 
         \hline
    \end{tabular}
    \caption{Experimental conditions for single diaphragm experiments on HPST. Measured values of the incident shock velocity and pressures computed from ideal shock relations are also tabulated.}
    \label{tab:Experimental conditions}
    \end{center}
\end{table*}

\begin{table}
    \def~{\hphantom{0}}
    \begin{center}
    \begin{tabular}{ccc}
        \# & Distance from diaphragm& Normalized distance\\[3pt]
 & in m&$X/D$\\
        \hline
        1  & 0.19 & 1.89 \\
        2  & 0.67 & 6.64 \\
        3  & 1.15 & 11.39 \\
        4  & 1.51 & 14.89 \\
        5  & 1.99 & 19.64 \\
        6  & 2.47 & 24.39 \\
        7  & 2.94 & 28.96 \\
        8  & 3.73 & 36.71 \\
        9  & 4.51 & 44.46 \\
        10 & 5.28 & 52.01 \\
        11 & 6.13 & 60.36 \\
        12 & 6.59 & 64.88 \\
        \hline
    \end{tabular}
    \caption{Axial locations of the pressure transducers mounted on the driven section of the shock tube.}
    \label{tab:PCB_location}
    \end{center}
\end{table}

\subsection{Diaphragm opening dynamics}
\begin{figure*}
    \centering
    \includegraphics[width=0.75\textwidth]{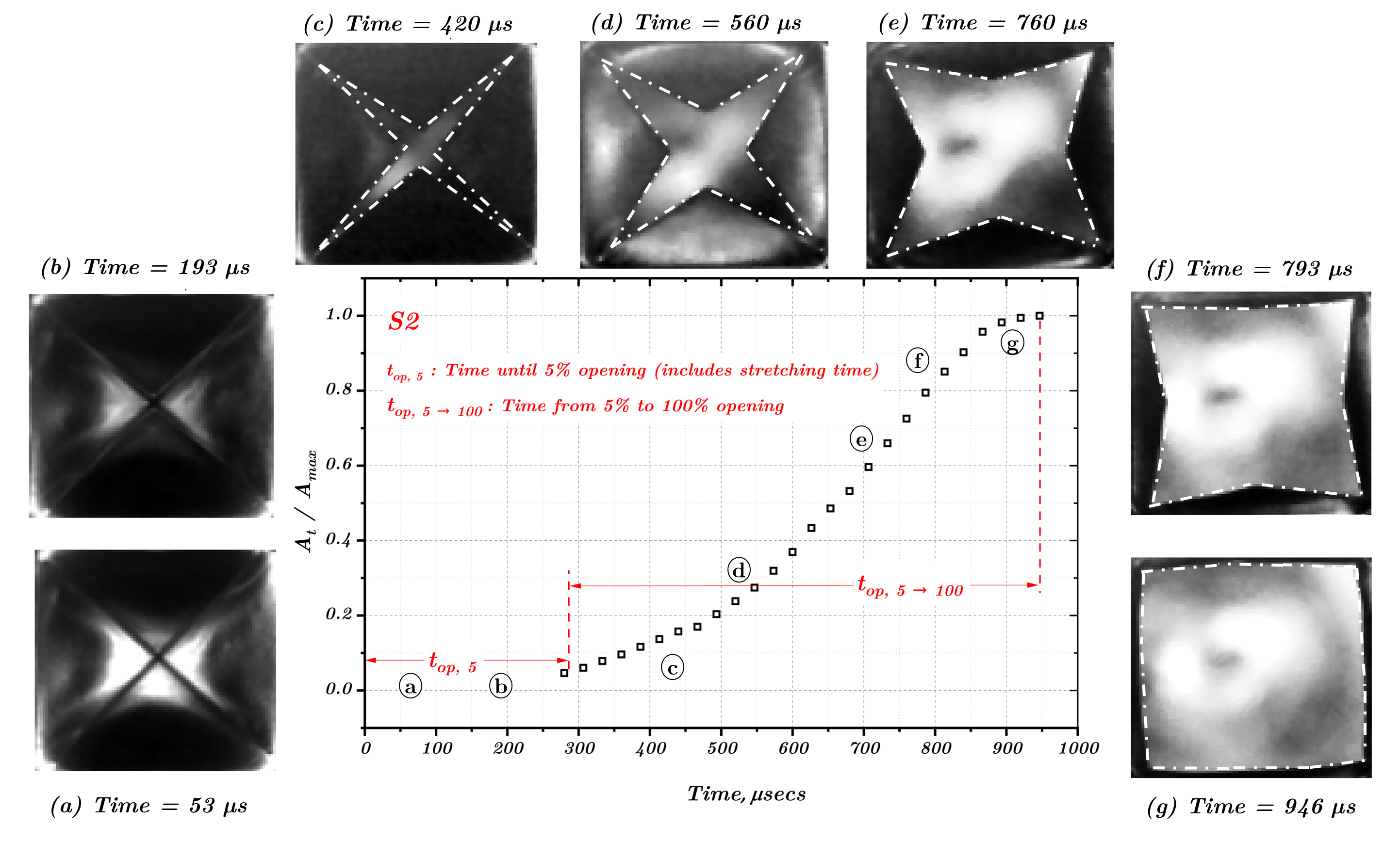}
    \caption{The diaphragm opening process as visualized using a high-speed camera placed at the end wall for case S2. The line plot shows the opened area as a function of time. The time zero indicates the instant when the diaphragm stretching began. Scatter points indicate the opened aperture area measured after 5\% of opening. Multimedia view shows the diaphragm opening as a function of time for all the seven cases. (Multimedia available online)} 
    \label{fig:Diaphragm_opening}
\end{figure*}

The images of the diaphragm rupture recorded at various stages for a typical experiment are shown in Fig. \ref{fig:Diaphragm_opening}. The diaphragm opening process can be divided into multiple stages. The first stage comprises of the stretching phase in which the diaphragm undergoes plastic deformation before reaching its rupture threshold due to the differential pressure between the driver and driven sections of the tube. In Fig. \ref{fig:Diaphragm_opening}, the image \textit{a} corresponds to this stretching phase. The bright white region in the center is due to the reflection from the light source, placed near the end wall as shown in the setup schematic (Fig. \ref{fig:Schematic_shock_tube}). For a more detailed interpretation of the visualization of the diaphragm opening process, readers are referred to Fig. \ref{fig:Diaphragm_opening} (Multimedia view). Following the initial stretching phase, the initial rupture of the diaphragm typically forms at its center, where the stress concentration is maximum. The rupture then propagates along predefined lines of weakness (the 'X' pattern scored during diaphragm manufacturing) to guide the tear in a controlled and symmetrical manner. Image \textit{b} in the Fig. \ref{fig:Diaphragm_opening} corresponds to this stage. Due to insufficient contrast (in the image \textit{b}) between the petal and the opened aperture, there is limited information on the extent of the tear along the lines of weakness making it challenging to track the edges of the tear to quantify the opened area during this phase. Tracking the edges becomes feasible after the opened area reaches approximately  $\simeq$ 5$\%$ of the final aperture (tracked edges shown as white dashed lines in images \textit{c} - \textit{g}). The scattered points in the plot in Fig. \ref{fig:Diaphragm_opening} signify the normalized area of diaphragm opening as a function of time, post 5$\%$ opening. The initial stretching and tearing process up to 5$\%$ can take hundreds of microseconds, as tabulated in Table \ref{tab:Diaphragm opening} under the label $t_{op,5}$. After the diaphragm is fully ruptured along the lines of weakness, the bending of the petals becomes the dominant mechanism, which is a significantly rapid process. During this phase, the diaphragm petals undergo pure bending, characterized by large deflections in the petals of the diaphragm observed in the images \textit{c, d, e,} and \textit{f}. As the petals approach the shock tube walls, the buildup of pressure between the petal and wall offers resistance to the petal motion and slows the bending process. Furthermore, there is a reduction in the driver pressure acting on the petal surface as the driver gas expands into the driven section. Therefore, the final stages of the opening process sees a significant deceleration in the opening process. 

\begin{table*}
    \def~{\hphantom{0}}
    \begin{center}
    \begin{tabular}{ccccccc} 
         Notation&  $P_{41}$ (-)&  $th_{eff}$ (mm) &  $t_{op,5}$ ($\mu$s) & $t_{op,100}$ ($\mu$s) & $\frac{t_{op,5}}{t_{op,100}}$ (-) & $t_{op,5 \rightarrow 100}$ ($\mu$s) \\ [3pt]
         \hline
         S1&  70&  0.50&   293& 998& 0.29 &  705 \\ 
         S2&  75&  0.59&   286& 946& 0.30 & 660 \\ 
         S3&  95&  0.68&   280& 800& 0.35 & 520 \\ 
         S4&  118&  0.72&   274& 700& 0.39 & 426 \\ 
         S5&  133&  0.76&   267& 672& 0.39 & 405 \\ 
         S6&  151&  0.81&   252& 612& 0.41 & 360 \\ 
         S7&  191&  1.00&   226& 546& 0.41 & 320 \\ 
         \hline
    \end{tabular}
    \caption{Diaphragm opening times including effective thickness and opening times.}
    \label{tab:Diaphragm opening}
    \end{center}
\end{table*}

In most cases, the diaphragm opening remains symmetric, as seen in the Fig. \ref{fig:Diaphragm_opening}(Multimedia view). The movie shows the opening process of the diaphragm and a red tracer tracks the opened area in the line plot in Fig. \ref{fig:Diaphragm_opening}(Multimedia view). The time taken for the diaphragm to open from 5$\%$ to 100$\%$ aperture area is referred to as $t_{op,5\rightarrow100}$ and the total opening time is termed as $t_{op,100}$. $t_{op,100}$ can be expressed as a sum of $t_{op,5}$ and $t_{op,5 \rightarrow 100}$, as shown in Eq. \ref{eq:t_100}.

\begin{equation}
t_{op,100} = t_{op,5} + t_{op,5 \rightarrow 100}   
\label{eq:t_100}
\end{equation}

\begin{figure*}
    \centering
    \includegraphics[width=0.9\textwidth]{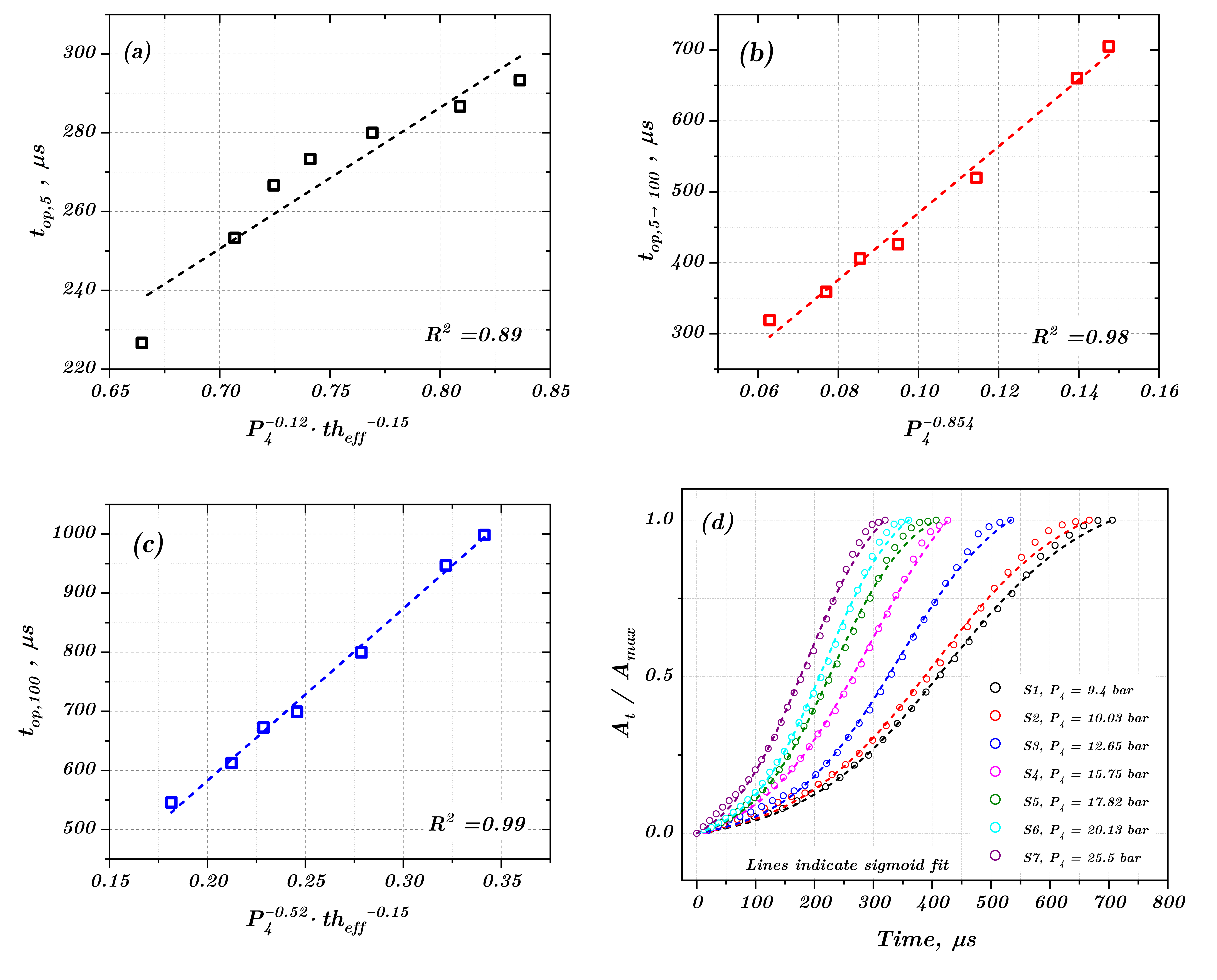}
    \caption{Experimentally measured values of (a) $t_{op,5}$, (b) $t_{op,5 \rightarrow 100}$, and (c) $t_{op,100}$ along with their respective fitted curves, and (d) Diaphragm opening profile measurements with a sigmoid fit (\ref{eq:sigmoid}) to represent the opening behavior.}
    \label{fig:Diaphragm_sigmoid_fit}
\end{figure*}

The values of the three opening times observed in the experiments are listed in Table \ref{tab:Diaphragm opening}. Previous studies \cite{white1958influence,rothkopf1974diaphragm} show that $t_{op,5}$ can take about 20 - 25$\%$ of the total opening time, $t_{op,100}$. However, in the present experiments, it is seen that $t_{op,5}$ can go up to 41$\%$ of the total opening time at higher driver pressures (Table \ref{tab:Diaphragm opening}). There are no correlations for $t_{op,5}$ in literature to the best of authors' knowledge. If $t_{op,5}$ is assumed to be dependent on the driver pressure, $P_{4}$, and the effective thickness of the diaphragm at the notch, $th_{eff}$, then a correlation of the form (Eq. \ref{eq:Correlation_5_percent}) can be formulated,  

\begin{equation}
t_{op,5}, \mu s = f(P_4,th_{\text{eff}}) = a_1 \cdot (P_4)^{a_2} \cdot (th_{\text{eff}})^{a_3} 
\label{eq:Correlation_5_percent}
\end{equation}

where $t_{op,5}$ is in $\small{(\mu s)}$, $P_{4}$ is in bar, $th_{eff}$ is in mm, and $a_1$, $a_2$, $a_3$ are constants. Fitting this correlation to the present data points yields $a_1=358$, $a_2=-0.12$, and $a_3=-0.15$. The agreement with the correlation is good as shown in Fig. \ref{fig:Diaphragm_sigmoid_fit}a. 

The opening time from 5\% to 100\% opening $(t_{op,5 \rightarrow 100})$, is predominantly a bending process for the diaphragm, dependent on the diaphragm thickness and driver pressure ($P_4$). Across all the cases the thickness at the bending location is the same, hence a correlation for $(t_{op,5 \rightarrow 100})$ is developed with dependence on only $P_4$ (Eq. \ref{eq:Correlation_5_to_100_percent}), as shown in Fig. \ref{fig:Diaphragm_sigmoid_fit}b. The resulting fit parameters are $b_1 = 4700$, $b_2 = -0.854$.

\begin{equation}
t_{op,5 \rightarrow 100}, \mu s  = f(P_4) = b_1 \cdot (P_4)^{b_2}  
\label{eq:Correlation_5_to_100_percent}
\end{equation}

In contrast, $t_{op,100}$, encompasses all three stages of diaphragm opening (stretching, tearing, and bending). Hence any correlation developed for $t_{op,100}$ must be dependent on the $th_{eff}$ and $P_4$. The correlation is given in Eq. \ref{eq:Correlation_100_percent} with fit parameters $c_1 = 2913$, $c_2 = -0.52$, and $c_3 = -0.15$ and predictions shown in Figure \ref{fig:Diaphragm_sigmoid_fit}c.

\begin{equation}
t_{op,100}, \mu s  = f(P_4,th_{\text{eff}}) = c_1 \cdot (P_4)^{c_2} \cdot (th_{\text{eff}})^{c_3} 
\label{eq:Correlation_100_percent}
\end{equation}

The exponent for $P_4$ turns out to be -0.52, this closely matches the inverse square root dependence showed by Drewry et al.\cite{drewry1965determination} for total opening times of diaphragms.


In addition to these correlations, a sigmoidal function was used to model the evolution of the opened aperture area as a function of time. The following sigmoid equation (Eq. \ref{eq:sigmoid}) was fitted to the experimental data for each case, and the predictions from the correlations are shown in Figure \ref{fig:Diaphragm_sigmoid_fit}d. 

\begin{equation}
\frac{A_t}{A_{max}} = rescale\left[ \frac{1}{1 + exp\left( \eta ( \frac{t}{t_{op,5 \rightarrow 100}} - \kappa) \right)} \right]
\label{eq:sigmoid}
\end{equation}

where $A_t$ is the opened aperture area until time $t$, $A_{max}$ is the total area. The \textit{rescale} function in MATLAB which normalizes the area between $0$ and $1$ was also included in the objective function for fitting the variables $\eta$ and $\kappa$. The fitted parameter $\eta$ lies in the range $\eta \in [4.75, 5.7]$, and $\kappa$ lies in the range $\kappa \in [-0.68 -0.57]$. The sigmoidal fit captures the gradual acceleration of the diaphragm opening, followed by the rapid opening phase and subsequent deceleration, as described earlier. Fig. \ref{fig:Diaphragm_sigmoid_fit}d shows that this function provides a good match to the experimentally measured opening areas at various time intervals. These fitted correlations and the sigmoidal function have been used to model the diaphragm opening process in the CFD simulations of the shock tube, ensuring that the opening dynamics are captured accurately throughout the simulation.

\section {Numerical model of shock tube}

\begin{figure*}
    \centering
    \includegraphics[width=0.9\textwidth]{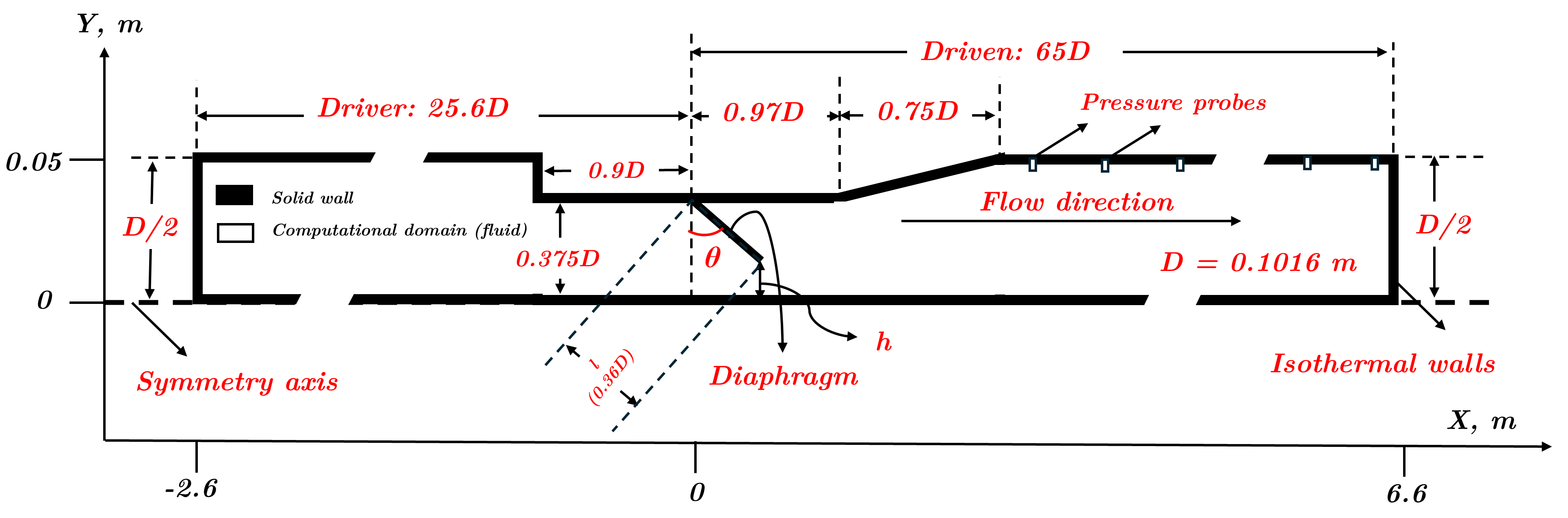}
    \caption{Schematic describing the numerical domain for the shock tube including the incorporation of diaphragm opening in CFD.} 
    \label{fig:Numerical_domain}
\end{figure*}

\subsection {Model setup}
The numerical domain consists of a 2.6 m driver section and a 6.6 m driven section, with only half of the shock tube being modeled assuming symmetry about the center line (shown in Fig. \ref{fig:Numerical_domain}). It should be noted that the simulation encompasses only the regular length of 6.6 m of the driven section, excluding the additional length equipped with optical diagnostics due to the absence of pressure transducers in that section. The shock tube walls are treated as isothermal, with a wall temperature of $T_w = 294.6$ K. The initial pressure in the driver and driven sections are based on the experimental values provided in Table \ref{tab:Experimental conditions}. Both the driver and driven gases are initialized a temperature of 294.6 K. The diaphragm opening process is modeled as a petal hinged about a point on the shock tube wall at the diaphragm location (shown in Fig. \ref{fig:Numerical_domain}). As the petal rotates anticlockwise from the initially closed position wall it gradually increases the area of the orifice between the driver and the driven section. The length of the diaphragm petal, $(l)$, is taken as 36.5 mm. At time zero, the orifice height between the driver and driven sections is initially 0.2 mm and the opening of the diaphragm is governed by Eq. \ref{eq:height_function}. 
\begin{equation}
h = 
\begin{cases}
    0.2 \, mm & \text{if } t = t_0 \\
    0.2 + \ell(1 - \cos \theta), mm & \text{if } \hspace{1mm} t_0 < t \hspace{1mm} \leq \hspace{1mm} t_{op}^{exp}  \\
    0.2 + \ell \ , mm & \text{if } t \hspace{1mm} > \hspace{1mm} t_{op}^{exp}
\end{cases}
\label{eq:height_function}
\end{equation}
where $t_{op}^{exp}$ is the diaphragm opening time obtained through imaging experiments and $t_0$ corresponds to time zero. Eq. \ref{eq:height_function} controls the opening from $t_0$ until the experimentally measured diaphragm opening time until 100\% opened area. The value of $\theta$ in Eq. \ref{eq:height_function} is given by Eq. \ref{eq:theta}.

\begin{equation}
\theta = \cos^{-1} \left(1 - \frac{A_t}{A_{max}}\right)
\label{eq:theta}
\end{equation}

where $\frac{A(t)}{A_{max}}$ is the normalized experimentally measured diaphragm opening profile, the same term described before in Eq. \ref{eq:sigmoid}. At $t > t_{op}^{exp}$, the diaphragm remains fully open. 

\subsection{Governing equations and solver settings}

The two-dimensional simulations were performed using a commercial CFD software, CONVERGE CFD (v3.1)\citep{converge2024}. The two-dimensional (2D) transient Navier–Stokes equations can be expressed as follows.

\begin{equation}\label{eq:continuity}
    \frac{\partial \rho}{\partial t} + \frac{\partial (\rho u_j)}{\partial x_j} = 0
\end{equation}

\begin{equation}\label{eq:momentum}
    \frac{\partial (\rho u_i)}{\partial t} + \frac{\partial (\rho u_i u_j)}{\partial x_j} = - \frac{\partial p}{\partial x_i} + \frac{\partial \hat{\tau}_{ji}}{\partial x_j}
\end{equation}

\begin{equation}\label{eq:energy}
    \frac{\partial (\rho E)}{\partial t} + \frac{\partial (\rho u_j H)}{\partial x_j} = \frac{\partial}{\partial x_j} \left( u_i \hat{\tau}_{ij} + (\mu + \sigma^* \mu_T) \frac{\partial k}{\partial x_j} - q_j \right)
\end{equation}

where $\rho$ is the density, $u$ is the velocity, $p$ is the pressure, $E$ is the total specific energy, $\mu$ is the molecular viscosity (Pa·s), $\hat{\tau}_{ij}$ is the shear stress tensor, $H$ is the total enthalpy, $k$ is the specific turbulence kinetic energy.

\begin{equation}\label{eq:stress_tensor}
    \hat{\tau}_{ij} = (\mu + \mu_T) \left( \frac{\partial u_i}{\partial x_j} + \frac{\partial u_j}{\partial x_i} - \frac{2}{3} \frac{\partial u_k}{\partial x_k} \delta_{ij} \right) - \frac{2}{3} \rho k \delta_{ij}
\end{equation}

The total specific energy and enthalpy are given by:

\begin{equation}\label{eq:energy_enthalpy}
    E = (C_v T + u_i u_i / 2 + k), \quad H = (C_p T + u_i u_i / 2 + k)
\end{equation}

The transport equations for the specific turbulence kinetic energy $k$ and dissipation rate $\omega$ are:

\begin{equation}\label{eq:k_transport}
    \frac{\partial (\rho k)}{\partial t} + \frac{\partial (\rho u_j k)}{\partial x_j} = P_k (\beta^* \rho \alpha k) + \frac{\partial}{\partial x_j} \left[ (\mu + \sigma_k \mu_T) \frac{\partial k}{\partial x_j} \right]
\end{equation}

\begin{subequations}\label{eq:omega_transport_split}
\begin{align}
    \frac{\partial (\rho \omega)}{\partial t} + \frac{\partial (\rho u_j \omega)}{\partial x_j} &= P_k (\beta \rho \omega^2) + \frac{\partial}{\partial x_j} \left[ (\mu + \sigma_\omega \mu_T) \frac{\partial \omega}{\partial x_j} \right] \\
    &+ 2 \rho (1 - F_1) \sigma_{\omega_2} \frac{1}{\omega} \frac{\partial k}{\partial x_j} \frac{\partial \omega}{\partial x_j}
\end{align}
\end{subequations}

where $\omega$ is the specific dissipation rate. The constants $C_v$ and $C_p$ denote the specific heat capacities at constant volume and pressure, while $Pr$ is the Prandtl number. The function $F_1$ represents the blending function in the $k-\omega$ model. where $F_1$ is the blending function between the standard $k-\omega$ model ($F_1 \to 1$ near the wall) and the $k-\epsilon$ model ($F_1 \to 0$). The eddy viscosity is given by:

\begin{equation}\label{eq:eddy_viscosity}
    \mu_T = \alpha^* Re_T \mu
\end{equation}

The net production per unit dissipation for $k$ and $\omega$ is expressed as:

\begin{equation}\label{eq:pk}
    P_k = \frac{\alpha^*}{\beta^*} \left( \tau_{ij} \frac{\partial u_j}{\partial x_j} \frac{1}{\mu_T \omega^2} \right) - 1
\end{equation}

\begin{equation}\label{eq:pomega}
    P_\omega = \frac{\alpha \alpha^*}{\beta} \left( \tau_{ij} \frac{\partial u_j}{\partial x_j} \frac{1}{\mu_T \omega^2} \right) - 1
\end{equation}

The parameters $\alpha, \alpha^*, \beta, \beta^*$ are $k-\omega$ closure coefficients, as are $\sigma_k, \sigma_\omega, \sigma^*$.

\begin{figure*}
    \centering
    \includegraphics[width=0.99\textwidth]{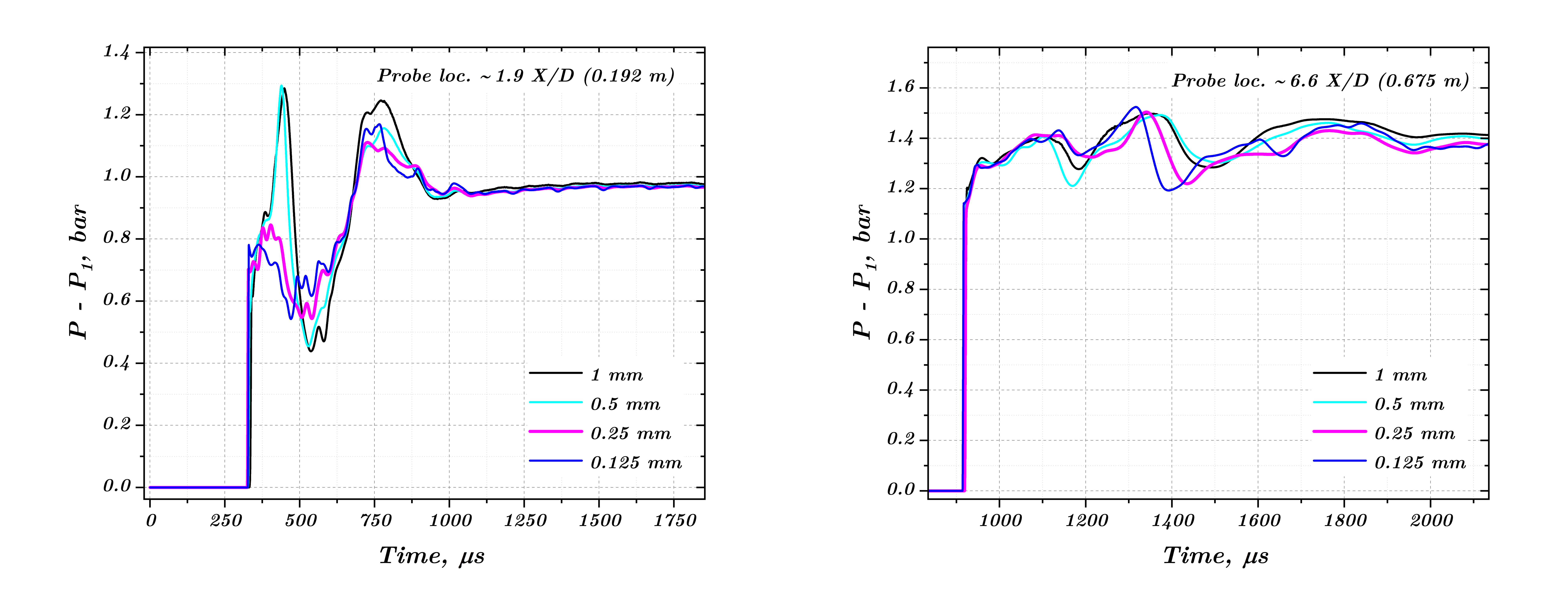}
    \caption{Grid convergence study comparing the pressure measurements at two locations (1.9 and 6.6 X/D) for minimum grid sizes from 1 - 0.125 mm for case S1. The recommended grid sizes based on this study is 0.25 mm} 
    \label{fig:Grid_convergence}
\end{figure*}

A density-based solver was employed with the PISO algorithm (Pressure-Implicit with Splitting of Operators) for pressure-velocity coupling. Spatial discretization was handled using the Monotonic Upstream-Centered Scheme for Conservation Laws (MUSCL) to achieve third-order accuracy. Temporal discretization was implemented using an adaptive time-stepping method, where the time step was limited by the diaphragm motion speed and a Mach-based CFL condition to maintain numerical stability ranging between (5e-9s to 9e-8s). CONVERGE CFD utilizes a cartesian cut-cell method for mesh generation at every time step to accommodate the moving diaphragm boundaries. A base mesh size of 2 mm was used with adaptive mesh refinement applied to refine the cells locally based on the temperature-, pressure-, and velocity-difference criteria between the neighboring cells, which is effective in reducing the overall simulation time. Three adaptive mesh refinement levels were used, refining the cells to a size of 0.25 mm (minimum cell size $\small{\frac{1}{4}}$ base grid/$2^3$). This minimum grid size was selected based on a grid dependency study with varying minimum grid sizes from 1 mm to 0.125 mm, as shown in Fig. \ref{fig:Grid_convergence}.

The simulations also included grid embedding around critical zones like the diaphragm rupture area to improve accuracy without excessive computational cost. The $k$-$\omega$ SST turbulence model (Eqs. \ref{eq:k_transport} - \ref{eq:pomega}), which provides an accurate representation of near-wall turbulence and flow separation while maintaining robustness for shock-driven flows, was utilized. Wall heat transfer was modeled using the Han and Reitz model, which governs thermal exchange based on convective heat transfer correlations. The Law of the Wall model was applied for both velocity and temperature boundary conditions. A surface roughness parameter was included to account for the influence of near-wall interactions on flow evolution. This setup contributed to improved agreement with expected shock attenuation trends, likely due to enhanced boundary layer effects and thermal energy dissipation.

\subsection{Model validation}
\begin{figure*}
    \centering
    \includegraphics[width=0.7\textwidth,height=0.6\textheight]{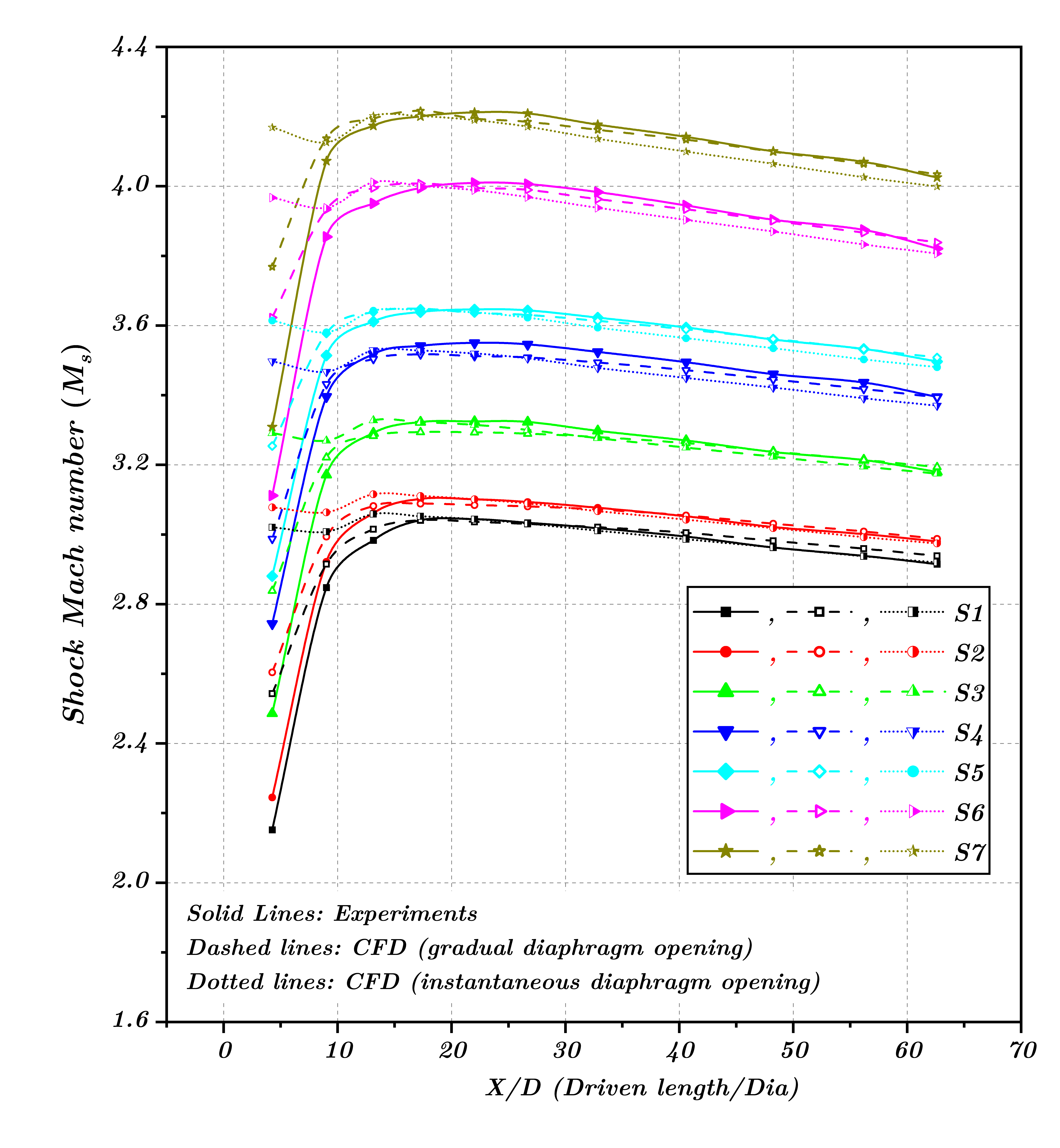}
    \caption{Comparison of experimental and CFD shock velocity profiles for all test cases. Solid markers represent experimental data, while dashed lines with hollow symbols correspond to CFD predictions incorporating diaphragm opening. Half filled symbols connected with a dotted line denote CFD predictions with an instantaneous opening of the diaphragm.} 
    \label{fig:CFD_vs_EXP}
\end{figure*}

To evaluate the accuracy of the numerical simulations, a detailed comparison of the CFD and experimental measurements were performed between (i) the computed shock velocities at various axial locations along the shock tube and (ii) the pressure histories obtained at different sensor locations of the driven section. Previous studies\cite{ikui1969investigations1,kashif2024insights} have indicated that shock Mach number profiles in the single-diaphragm mode are typically characterized by an initial acceleration phase where the shock Mach number increases to a peak value followed by a deceleration phase when the boundary layer effects become dominant. The experimental results presented in Fig. \ref{fig:CFD_vs_EXP} (solid symbols) align with this well-established behavior. The scatter points represent the shock Mach number measured between successive pairs of pressure transducers, which provides a shock Mach number profile along the length of the shock tube. With twelve transducers mounted along the tube, eleven discrete data points were obtained, and these points were plotted at the midpoint between each pair of transducers. The scattered data are interconnected using a modified Bezier curve (solid lines in Fig. \ref{fig:CFD_vs_EXP}), created in Origin Pro software.

Across all driver pressures in Fig. \ref{fig:CFD_vs_EXP}, the shock profiles consistently exhibit an acceleration region up to about approximately 20 $X/D$ and a subsequent deceleration. Additionally, shock deceleration is more prominent at higher shock Mach numbers due to a thicker boundary layer behind the shock wave. The shock Mach number profile was obtained from the numerical simulations at monitor points located at the same axial distances where the pressure transducers were mounted in the experiments. Figure \ref{fig:CFD_vs_EXP} illustrates this comparison, where dashed lines with hollow symbols correspond to the CFD predictions with diaphragm opening incorporated from the experiments. The simulations show a fairly good agreement with the experimentally obtained shock acceleration and deceleration trends. The first CFD data point consistently over-predicts the shock Mach number in all cases which could be attributed to the numerical limitation as the model incorporates $(t_{op,5 \rightarrow 100})$ as the diaphragm opening time instead of $(t_{op,100})$ leading to the velocity profile shifted onto the left. Apart from this initial deviation, the peak velocities are captured well across all cases. Notably, the agreement between the experimental and simulated shock velocities during the deceleration phase is also good. Moreover, Fig. \ref{fig:CFD_vs_EXP} shows shock Mach numbers for obtained from CFD cases assuming an instantaneous opening of diaphragm (indicated by dotted lines and half filled symbols). It is clear that neglecting the diaphragm opening process in the numerical model does not adequately capture the shock acceleration phase, although the trend of shock wave attenuation is captured, albeit with higher attenuation values at the shock tube's end.

\begin{figure*}
    \centering
    \includegraphics[width=0.99\textwidth]{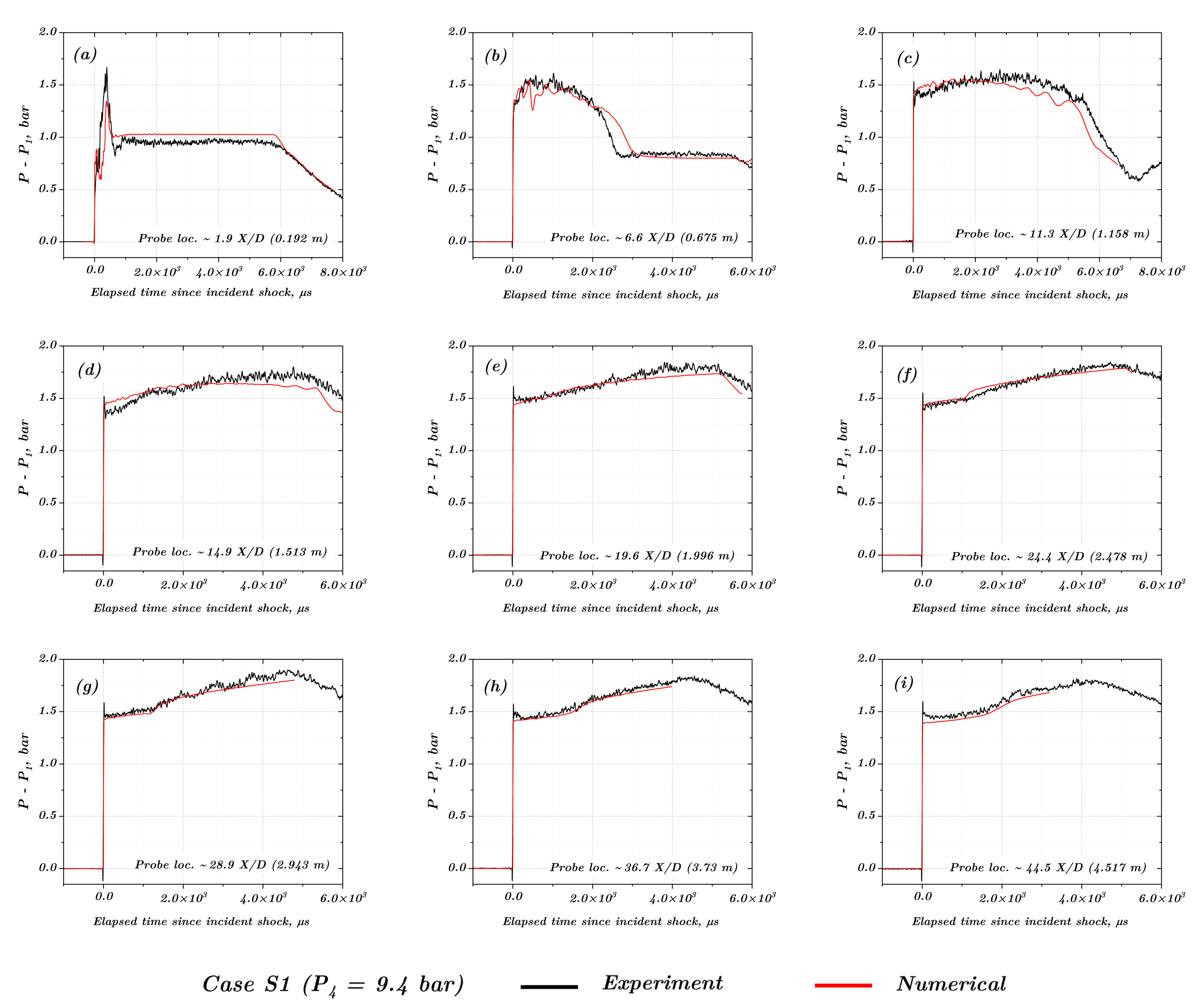}
    \caption{Comparison of experimental and CFD pressure profiles for case S1 for first 9 pressure transducers along the driven length. Black traces denote experimental measurements and red trace are for CFD measurements.} 
    \label{fig:Pressure_validation}
\end{figure*}

\begin{figure*}
    \centering
    \includegraphics[width=0.85\textwidth]{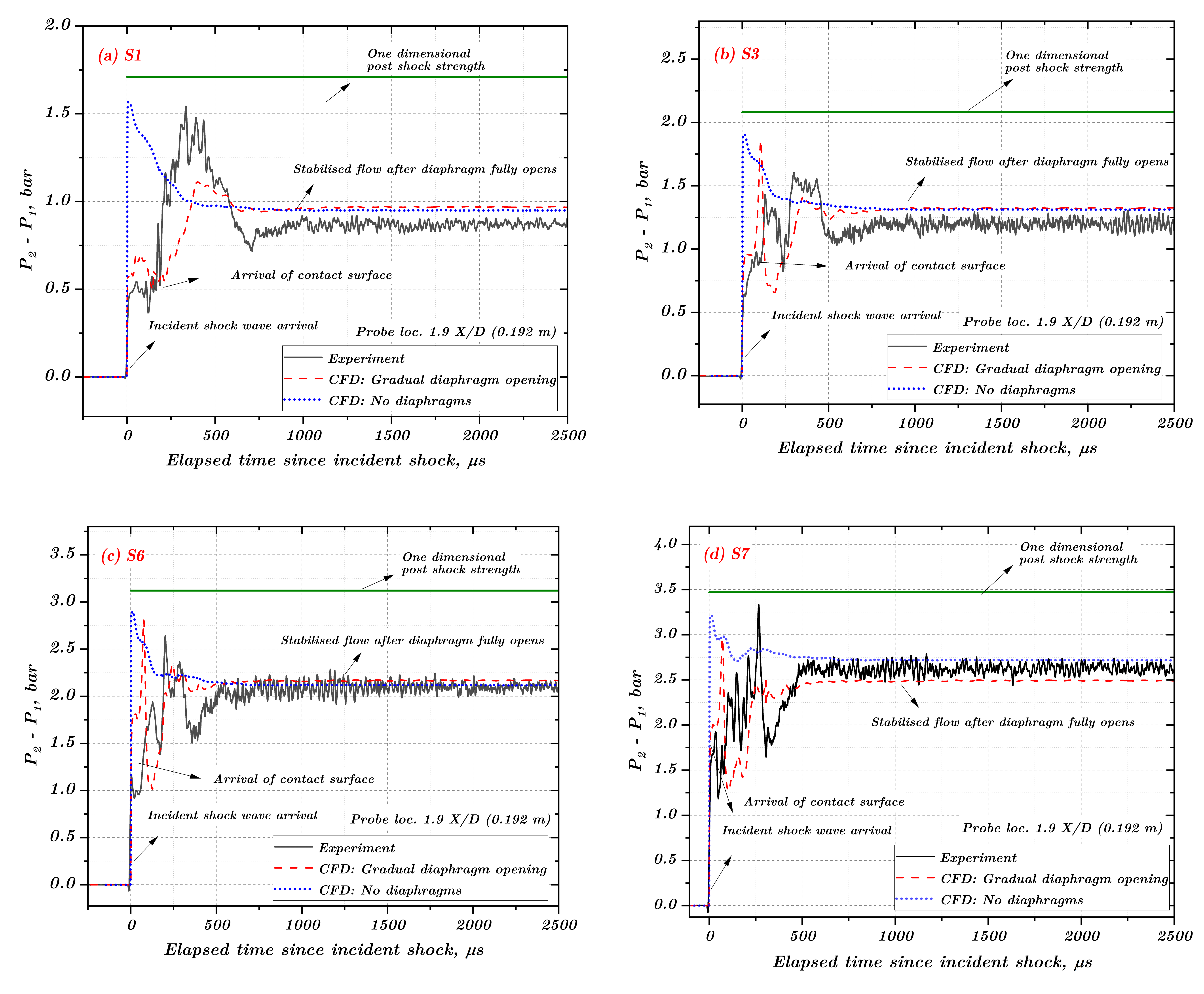}
    \caption{Comparison of pressure profiles from experimental measurements, CFD (with gradual opening of diaphragm), and CFD (with instantaneous diaphragm opening) for cases S1, S3, S6, and S7.} 
    \label{fig:S1_S7_First_PCB}
\end{figure*}

Further validation was done by comparing the pressure profiles at the different sensor locations along the driven section, as shown in Fig. \ref{fig:Pressure_validation} for case S1. The time instance of shock arrival in experiments and CFD is overlapped for better comparisons of the pressure profile. At $X/D = 1.9$ (Fig. \ref{fig:Pressure_validation}a), the simulations capture the magnitude of the initial pressure jump. Subsequently, a secondary pressure peak occurs, which is also qualitatively well-represented by the CFD results, though a slight reduction in the peak is observed in the simulation. Later in time, both the experimental and numerical traces exhibit similar stabilized pressure values, with the expansion wave arriving at comparable times. Without delving into each pressure comparison individually at higher $X/D$ values, it can be concluded that the CFD model shows good agreement with both the magnitude and profile of the pressure traces.

It must also be noted that the pressure profiles farther downstream ($X/D \geq 6.6$) show a consistent trend: a jump in pressure due to the shock wave, followed by a near constant-pressure region. The flow non-uniformities due to diaphragm opening are more visible at the first pressure transducer located at $X/D = 1.92$. The comparison of pressure profiles from CFD and experiments at this specific location is extended to other cases as well. Figure \ref{fig:S1_S7_First_PCB} shows such comparisons for cases S1, S3, S5, and S7. In addition to the CFD results with accurate diaphragm modeling, the figure also includes pressure profiles from CFD with instantaneous diaphragm opening and 1-dimensional post-shock strength (from ideal shock relations).

The time of arrival of the incident shock in the three modes—experiment, CFD with gradual diaphragm opening, and CFD with instantaneous opening—is again overlapped for better visualization. In case S1 (Fig. \ref{fig:S1_S7_First_PCB}a), the pressure rise due to the incident shock predicted by the CFD with gradual diaphragm opening (red trace) closely aligns with the experimental data (black trace). Shortly afterward, a subsequent pressure rise occurs in both traces, coinciding with the arrival of the contact surface (CS). In an ideal shock tube, the pressure across the CS remains uniform; however, this is not the case close to the diaphragm in real shock tubes, and this discrepancy is accurately captured in both the experimental and CFD results. Although the peak pressure is slightly under-predicted by the CFD, the pressure stabilizes and matches well with the experimental measurements. Figure \ref{fig:S1_S7_First_PCB}a also includes results from a numerical simulation assuming an instantaneous diaphragm opening (dotted blue line), which shows a linearly decreasing pressure and fails to capture the flow intricacies near the diaphragm. Similar conclusions hold true for cases S3 and S6 in Fig. \ref{fig:S1_S7_First_PCB}b and c, where the pressure oscillations are qualitatively well captured with the incorporation of diaphragm opening, while the instantaneous opening fails to predict the same.

In the highest driver pressure case, S7 (Fig. \ref{fig:S1_S7_First_PCB}d), the pressure trace comparisons reveal a slightly pronounced difference between the experimental and CFD results compared to the earlier cases. The initial rise in pressure due to the incident shock, as predicted by the CFD with gradual diaphragm opening (red trace), again aligns well with the experimental data (black trace). However, the pressure oscillations and rise corresponding to the arrival of the contact surface (CS) are not well captured by the simulations. This is likely due to the higher pressure in the driver section for case S7 causing a more rapid and turbulent expansion of the CS, resulting in larger fluctuations in the pressure trace and adding pronounced three-dimensional effects that are not captured by the 2-D simulations. Nonetheless, the qualitative pressures match, and the stabilized pressure is well predicted by the CFD. Overall, the numerical model with the implementation of diaphragm opening profiles from experiments provides a reliable framework that replicates the experimental measurements and can be utilized for further understanding the flow characteristics in the early stages of shock formation.

\section{Flow features due to the diaphragm rupture process}
The previous section demonstrated that the numerical model aligns well with the experimental measurements. Building on this, the current section delves deeper into analyzing the flow features near the diaphragm, which are challenging to capture experimentally. It investigates the wave system around the diaphragm station, the shape of the shock front, and the influence of the diaphragm opening process on the axial and radial non-uniformities in the shocked gas in proximity to the diaphragm. 

\subsection {Wave system near the diaphragm}
\begin{figure*}
    \centering
    \includegraphics[width=1\textwidth]{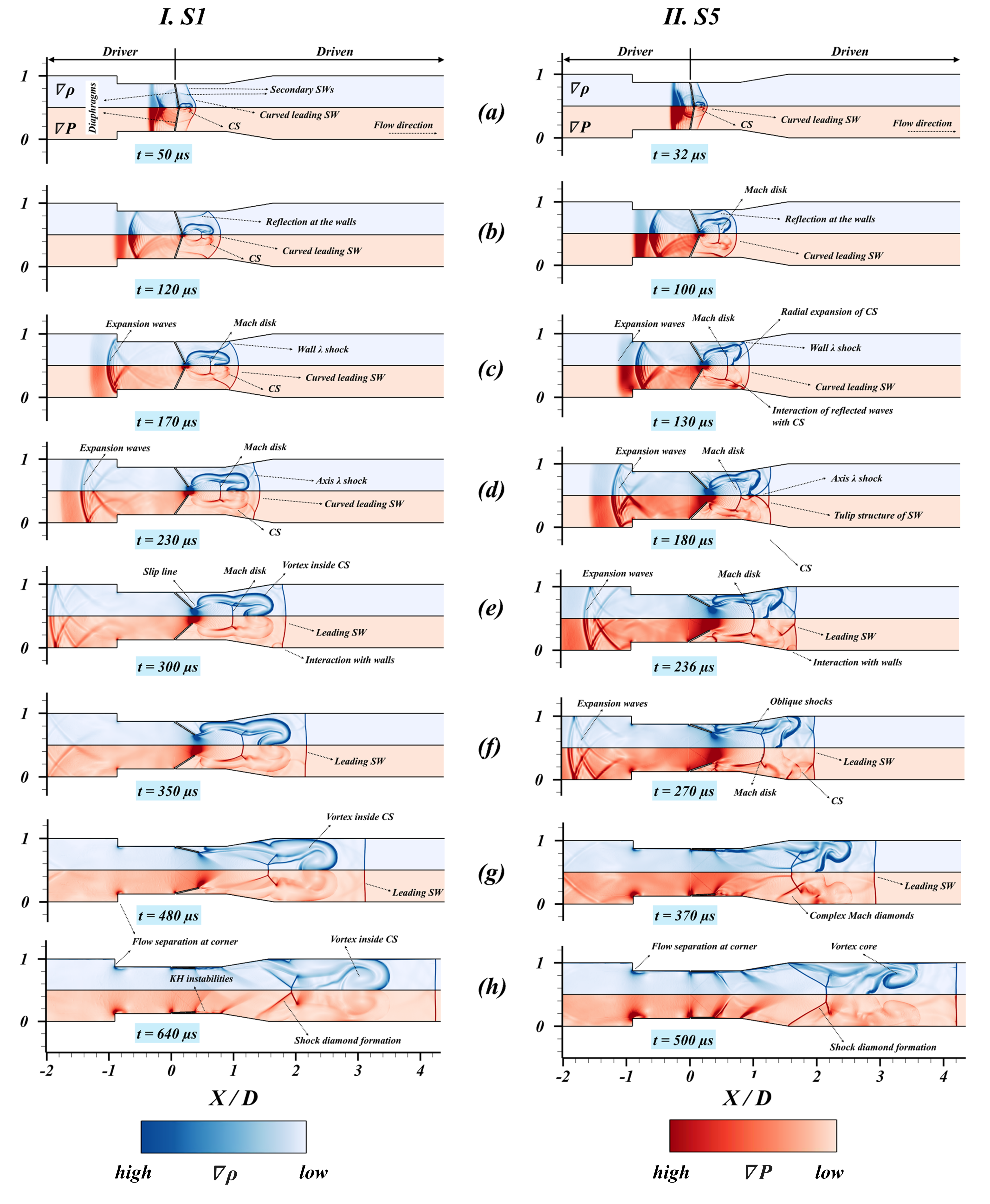}
    \caption{Gradient of density (top contour) and pressure (bottom contour) at various stages of diaphragm opening detailing the flow characteristics for cases S1 and S5. Multimedia extends these contours for all the cases. (Multimedia available online)} 
    \label{fig:S1_S5_flow}
\end{figure*}

As highlighted in the introduction, previous numerical studies on shock tubes have typically assumed non-rotating diaphragms, simulating an expanding aperture or iris instead.\cite{gaetani2008shock,currao2024diaphragm,petrie1998computational} However, this geometric assumption may limit the accuracy of these simulations in capturing the flow evolution near the diaphragm station. In contrast, the present work incorporates the rotational motion of diaphragm petals enabling a more realistic visualization of flow features. The flow characteristics within both the compressed driven gas (\textit{shocked}) and the expanding driver gas have been investigated. Notable differences are observed between the high and low diaphragm pressure-ratio cases. To illustrate these differences, density (top half) and pressure (bottom half) gradient contours are presented in Fig. \ref{fig:S1_S5_flow} for cases S1 and S5. Videos of the flow evolution for all cases are provided in Figure \ref{fig:S1_S5_flow} (Multimedia view).

\textbf{Low diaphragm pressure ratio (Case S1)}: The flow evolution for case S1 is depicted in Fig. \ref{fig:S1_S5_flow}I, highlighting complex shock structures and flow dynamics near the diaphragm at various stages of rupture. At t = 50 $\mu s$, a small aperture permits the driver gas to escape into the driven section, forming a curved leading shock wave that expands radially outward. Two secondary shock waves develop, attached to the leading shock wave, due to the rapid motion of the diaphragm and radial expansion of the contact surface (CS). Expansion waves propagate upstream in the driver gas, exhibiting slight curvature near the diaphragm. By t = 120 $\mu s$, the curved leading shock wave undergoes a regular reflection at the shock tube walls without forming Mach stems. The CS expands further developing into a strong discontinuity due to the expansion of the high-pressure driver gas.

At t = 170 $\mu s$, a Mach disk is formed in the driver gas which is a characteristic feature of under-expanded high-pressure gas. The shock interacts with shock tube walls, leading to the development of a wall $\lambda$-shock structure or Mach stem. By t = 230 $\mu s$, approximately 20\% of the cross-sectional area near the diaphragm is opened, and the leading shock propagates to 1.4 $X/D$. The shock front remains curved, with higher curvature near the center line of the driven section. An axisymmetric $\lambda$-shock structure becomes evident, similar to observations by Gaetani et al.\cite{gaetani2008shock} at low $P_{41}$ for instantaneous, yet partial, diaphragm openings.

At t = 300 $\mu s$, the leading shock wave exits the diverging section of the shock tube. Due to irregular reflection near the walls, reflected shock waves and Mach stems are produced. By t = 350 $\mu s$, the Mach disk within the driver gas extends to approximately 0.6 $D$, with two oblique shock waves attached to its edges. Multiple vortex cores are likely to form within the CS, although these are not clearly visible in the schlieren plots. By 2 $X/D$, the shock wave achieves near-planarity, and vortex cores seem to originate at slip lines near the diaphragm tips. At t = 640 $\mu s$, the leading shock wave becomes normal to the axis and propagates downstream with the CS. Oblique shock waves to the left of the Mach disk merge toward the center line. Kelvin-Helmholtz (KH) instabilities begin to form at the diaphragm tips, contributing to additional wave patterns as the flow expands, resembling under-expanded nozzle flows.

At later time steps, shown in Figure \ref{fig:S1_S5_flow} (Multimedia view), the classical shock diamond structure forms for all cases. The CS slowly assumes a parabolic shape. Due to the existence of a mixing length at the CS, the gradients across it become less sharp as the CS progresses downstream. Conversely, the \textit{shocked} driven gas does not exhibit strong wave patterns.

\textbf{High diaphragm pressure ratio (Case S5)}: Figure \ref{fig:S1_S5_flow}II illustrates contours for Case S5, highlighting key differences in flow behavior at higher pressures. At t = 32 $\mu s$, a small aperture allows the driver gas to expand, generating a curved leading shock wave that expands radially downstream. Two secondary shock waves are also formed, similar to Case S1. The expansion waves exhibit a steeper gradient compared to S1, attributed to the significantly higher $P_{41}$, as indicated by the darker regions upstream of the diaphragm. By t = 100 $\mu s$, the regular reflection of the curved shock wave against the walls produces reflected shock waves. A well-defined Mach disk is visible within the CS, and its width is significantly larger than that observed in S1.

At t = 130 $\mu s$, the reflected waves interact with the expanded CS, while the leading shock wave enters the expansion region, resulting in the irregular reflection of the shock wave and the formation of a $\lambda$-shock (Mach stem) near the walls. By t = 180 $\mu s$, an axisymmetric $\lambda$-shock forms near the center line, and the shock wave assumes a tulip structure, markedly different from S1. The shock front near the walls advances faster than along the center line due to the deformation of the CS, which exhibits faster radial expansion away from the tube's center line. By t = 236 $\mu s$, the leading shock wave is not yet fully planar as oblique shock waves continue to destabilize the main shock wave. The Mach disk begins to show signs of attached oblique shock formation.

At t = 270 $\mu s$, the shock wave progresses toward planarity as it moves downstream. By t = 370 $\mu s$, the leading shock wave becomes planar, although the CS remains highly deformed. At t = 500 $\mu s$, the diaphragm is fully open, and the planar leading shock wave propagates to approximately 4 $X/D$. The shock diamond feature begins to form as the driver gas under-expands. The contact surface exhibits multiple vortex cores visible in both density and pressure gradient contours. Similar to Case S1, as the shock wave moves downstream to 9 $X/D$ (refer to videos in Figure \ref{fig:S1_S5_flow} (Multimedia view)), the shocked driven gas does not exhibit steep discontinuities across all cases. However, mild variations in thermodynamic properties are observed in the shocked gas due to differences in shock velocity. These variations will be explored in detail in a subsequent section. 

\begin{figure}
    \centering
    \includegraphics[width=0.49\textwidth]{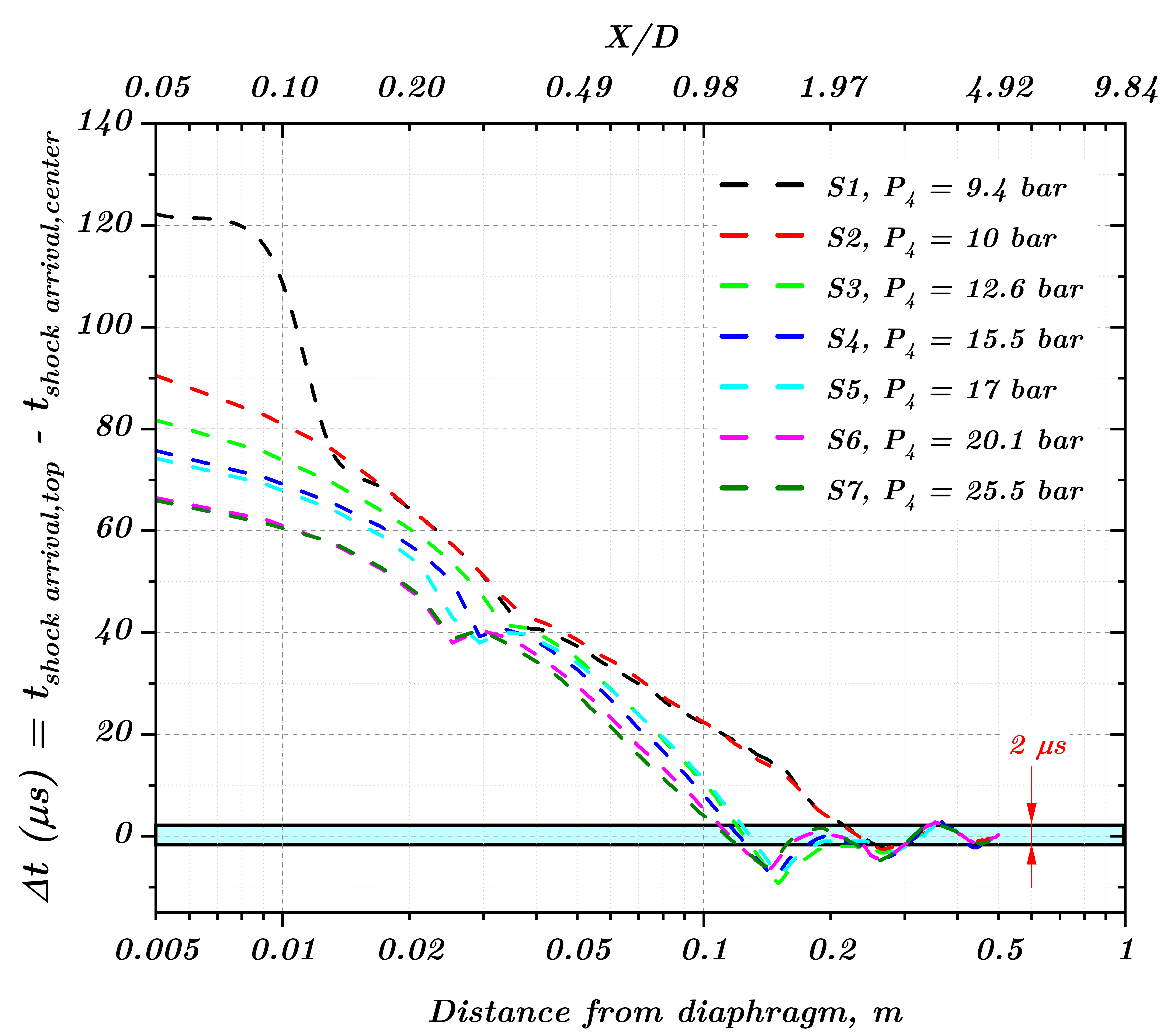}
    \caption{Shock planarity measurements and comparisons for cases S1 - S7.} 
    \label{fig:Shock_planarity}
\end{figure}

\textbf{Shock planarity}: In the flow contours presented in Fig. \ref{fig:S1_S5_flow}, the shock front is observed to achieve planarity (normal to the shock tube wall) at approximately 2 $X/D$ for case S1. In the case of S5, the shock front takes lesser time to achieve planarity indicating variations in the shock propagation characteristics. To precisely quantify the axial location when the shock front attains planarity, a new parameter $(\Delta t)$ is defined as the time difference between the shock arrival at the top and at the center-line of the shock tube. Figure \ref{fig:Shock_planarity} provides a visual representation of the $(\Delta t)$ measurements for all cases, with the \textit{x-axis} plotted on a logarithmic scale. A positive value of $(\Delta t)$ indicates that the shock arrives at the top location later than at the center-line, implying a concave shock wave structure, when viewed from the driven end. Conversely, a negative $(\Delta t)$ value signifies that the shock arrives at the top earlier than the center-line, resulting in a convex shock wave. The $(\Delta t)$ values are computed until they reach a threshold of $\leq 2 \mu s$, at which point the shock wave is considered to be planar.

Notably, for cases S1 and S2, shock planarity is achieved at a distance of approximately 0.2 m (or 2 $X/D$) from the diaphragm. Beyond this distance, $(\Delta t)$ begins to oscillate, but the amplitude of these oscillations remains less than 2 $\mu s$, indicating only minor deviations from planarity. In contrast, for cases S3 through S7, shock planarity is achieved at a shorter distance, within the range of 0.1 to 0.15 m from the diaphragm. However, the oscillations in $(\Delta t)$ continue over a greater length, extending up to approximately 0.3 m, suggesting that the shock front undergoes more fluctuations in these higher-pressure cases before reaching full stability. This methodology provides insight into the length scales over which shock waves attain planarity at different driver pressures. The variation in $(\Delta t)$ highlights the influence of pressure on shock propagation, with higher driver pressures leading to earlier shock planarity but more persistent oscillations. 

\subsection{Pressure and temperature variations in the shocked gas}
\begin{figure*}
    \centering
    \includegraphics[width=0.99\textwidth]{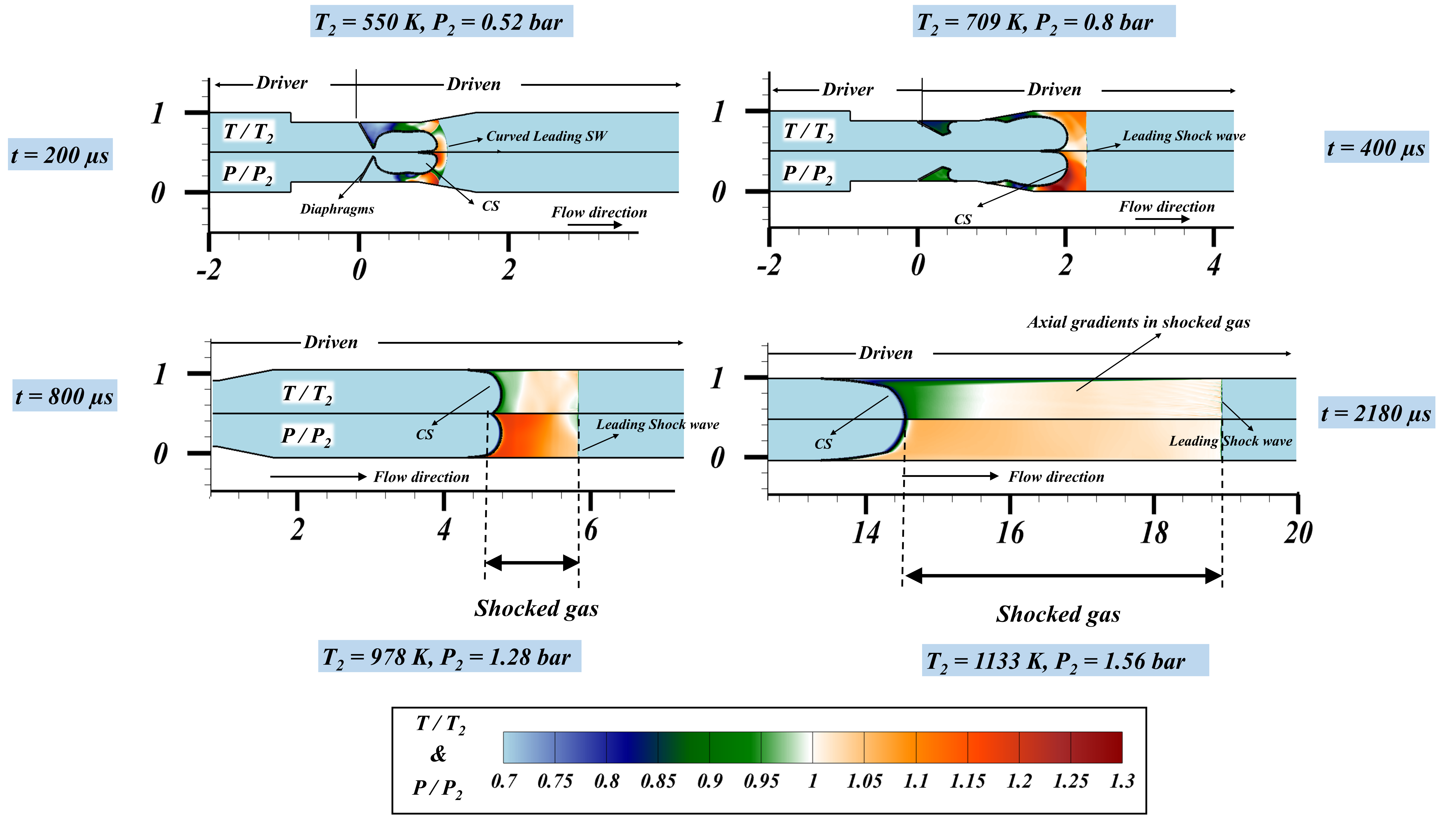}
    \caption{Contours depicting normalized temperature (top) and pressure (bottom) for case S2 at different time stamps. The reference values for normalization are acquired from instantaneous shock velocity and showed with the each of the panel. Multimedia view extends these contours for all the cases. (Multimedia available online)} 
    \label{fig:normalized_T_P}
\end{figure*}

The temperature and pressure inhomogeneities are significant during the shock formation stage, primarily due to (i) the continuous acceleration of the shock wave and (ii) the shock interactions with the shock tube wall. Temperature and pressure contours of the driven gas, normalized by the reference temperature and pressure values, are provided in Fig. \ref{fig:normalized_T_P}, and the multimedia view of Fig. \ref{fig:normalized_T_P} extends this all the cases. The reference values used for normalization are extracted from the local instantaneous shock velocity and ideal shock relations. The discussion on how instantaneous shock is obtained is provided in a later section (4.3). The justification for such normalization lies in the assumption that, at any given point in time, the thermodynamic conditions derived from the shock velocity represent the entire heated gas. These contours are useful for assessing the extent of variations in the thermodynamic properties across the entire heated slug compared to the gas immediately behind the shock wave.

From Fig. \ref{fig:normalized_T_P} (multimedia view), both radial and axial variations in temperature and pressure are evident up to 5 $X/D$. Pressure relaxation occurs rapidly when the local speed of sound is high due to elevated temperatures, yet the continuous variation of shock velocity maintains pressure non-uniformity up until 10 $X/D$. Once the shock has fully formed —indicated by the final frame in the video, where its acceleration has subsided—the axial variations in temperature and pressure become more pronounced. Radial variations in pressure are minimal, but temperature still exhibits radial variations due to the presence of cooler gas within the boundary layer. However, towards the end of the shock formation phase, the extent of differences in axial variations still exists for temperature, as is evident at t = 2180 $\mu$s in Fig. \ref{fig:normalized_T_P}.

\begin{figure*}
    \centering
    \includegraphics[width=0.99\textwidth]{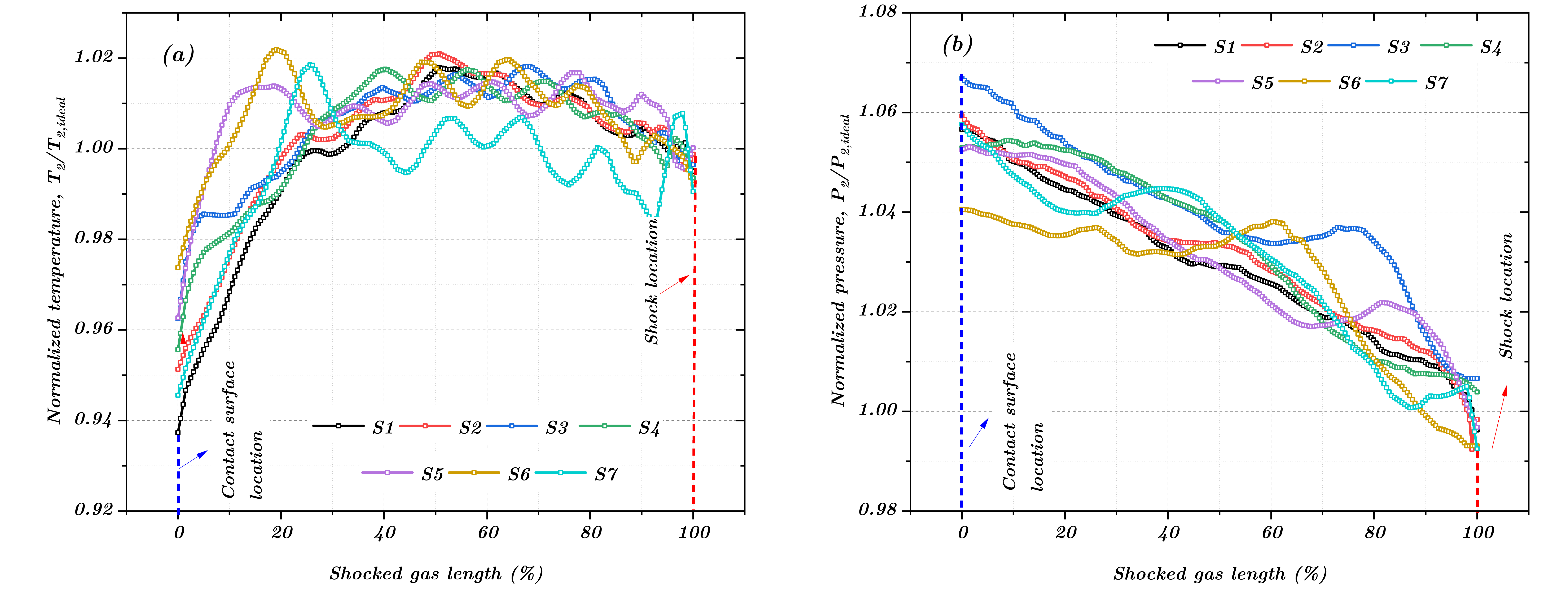}
    \caption{Axial distribution of temperature and pressure in the incident shocked gas. 0 \% on $x$-axis indicates the gas slug in front of the contact surface and 100 \% indicates the gas slug behind the shock.} 
    \label{fig:temperature_pressure_axial_variation}
\end{figure*}

The quantification of these axial gradients are given in the Figs. \ref{fig:temperature_pressure_axial_variation}a and b, where center-line normalized temperature and pressure and axial variations after the shock has stopped accelerating are shown. In this figure, 100\% on \textit{x-axis} denotes the shock location and 0\% denotes the CS, the region between the extremes represents the shocked gas. Normalization is done by temperature and pressure from the ideal shock relations taking the final shock velocity (after acceleration) as input. In Fig. \ref{fig:temperature_pressure_axial_variation}a, the temperature variation show a hotter region in the middle of the slug, and moving close the the contact surface the temperature drops, this could be due to the fact that this gas was processed the earliest with a lower shock velocity, the lowest temperature can be up to 6$\%$ less. Interestingly, the temperature profiles show a similar trend across all the cases, with slightly higher oscillations for case S7, could be due to more abrupt oscillation of shock velocity. Pressure axial variation on other hand (Fig. \ref{fig:temperature_pressure_axial_variation}b) show a linear increase in pressure moving away from the shock location, and a similar trend is observed across all the cases.

\subsection{Variations in shock Mach number}
\begin{figure}
    \centering
    \includegraphics[width=0.49\textwidth]{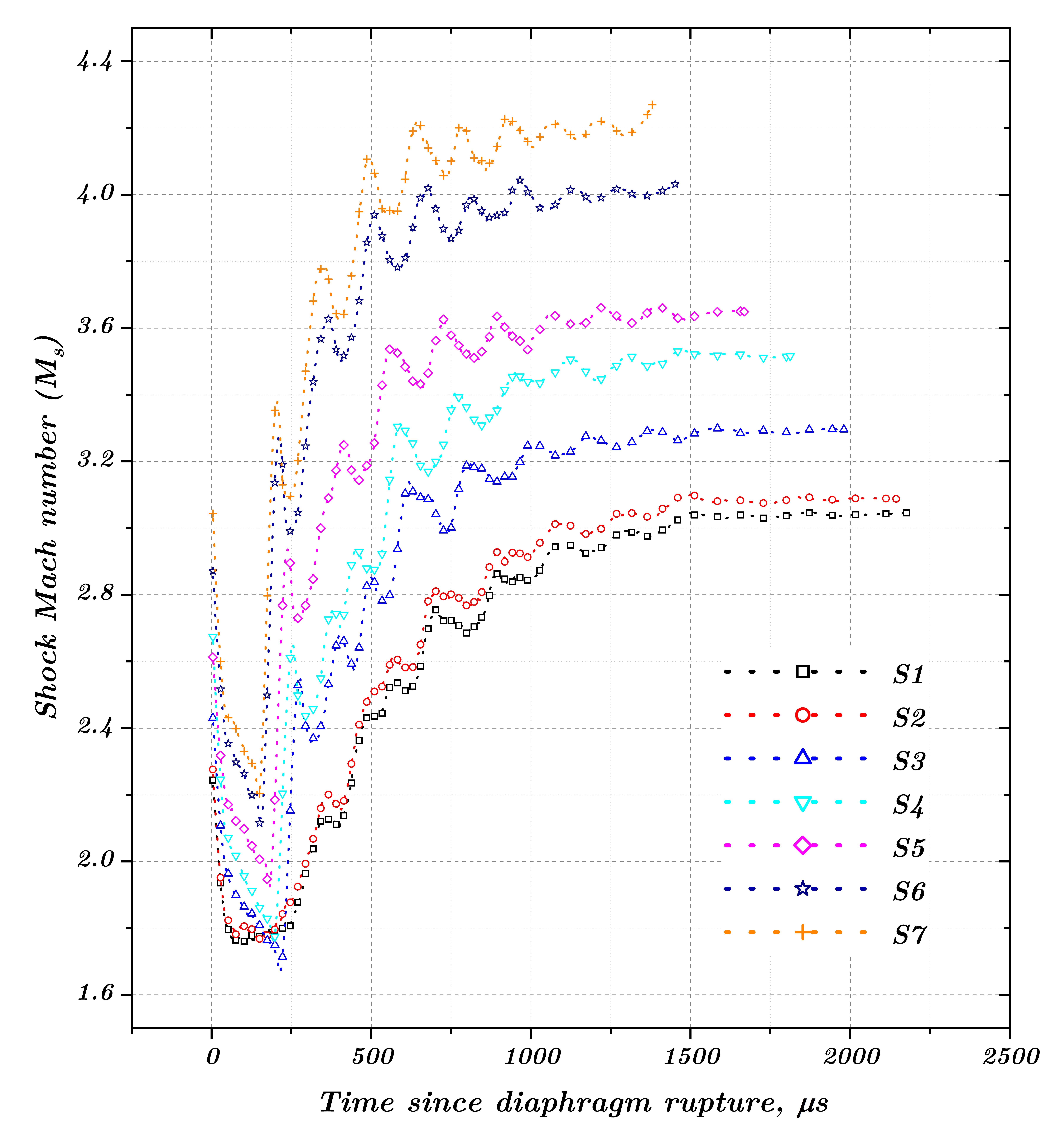}
    \caption{Instantaneous shock Mach number profiles from CFD (with incorporation of diaphragm opening) for cases S1 - S7, until the shock accelerates. The profiles have been smoothened with a Savitzky-Golay filter of window size 21.} 
    \label{fig:CFD_Instant_Shock_Vel}
\end{figure}

White\cite{white1958influence} proposed that the rupture of the diaphragm generates multiple compression waves that coalesce into a single discontinuity, which accelerates rapidly and transitions into a fully developed shock wave. Previous experimental studies have consistently reported shock acceleration during the initial phase, while numerical efforts that focus on validating these findings often neglect the diaphragm opening dynamics. Since experimental measurements of shock velocity are typically taken at discrete locations along the driven section, there is a spatial limitation that necessitates approximating the shock velocity profile near the diaphragm as a continuously accelerating trend, starting from Mach 1. However, numerical studies by Petrie-Repar\cite{petrie1998computational} and Alves et al.\cite{alves2021modeling} provide a more detailed temporal picture, revealing an initial deceleration phase during the early stages of shock evolution. This deceleration phase, often overlooked or misinterpreted as numerical artifacts or diaphragm modeling inaccuracies, warrant deeper investigation to better understand its physical origins. In the present study, a more accurate depiction of shock velocity variation is achieved by explicitly modeling the diaphragm's motion and opening. Numerical simulations (cases S1–S7) evaluate shock velocities in the first few meters of the driven section. By generating binarized temperature contours, the high-temperature shocked gas is distinguished from the low-temperature region 1 gas. A simple edge detection algorithm implemented in MATLAB continuously tracks the shock location with time. High-resolution contour images (0.2 mm pixel resolution) ensure precise shock front identification, improving fidelity in characterizing early shock dynamics. Figure \ref{fig:CFD_Instant_Shock_Vel} shows the variation in the shock Mach number at different operating conditions. The initial decrease in the shock Mach number followed by an increase to a peak value is noticeable in all cases. The profile of shock Mach number for a typical case can be generalized based on new parameters that are described in the following section.

\begin{figure*}
    \centering
    \includegraphics[width=0.9\textwidth]{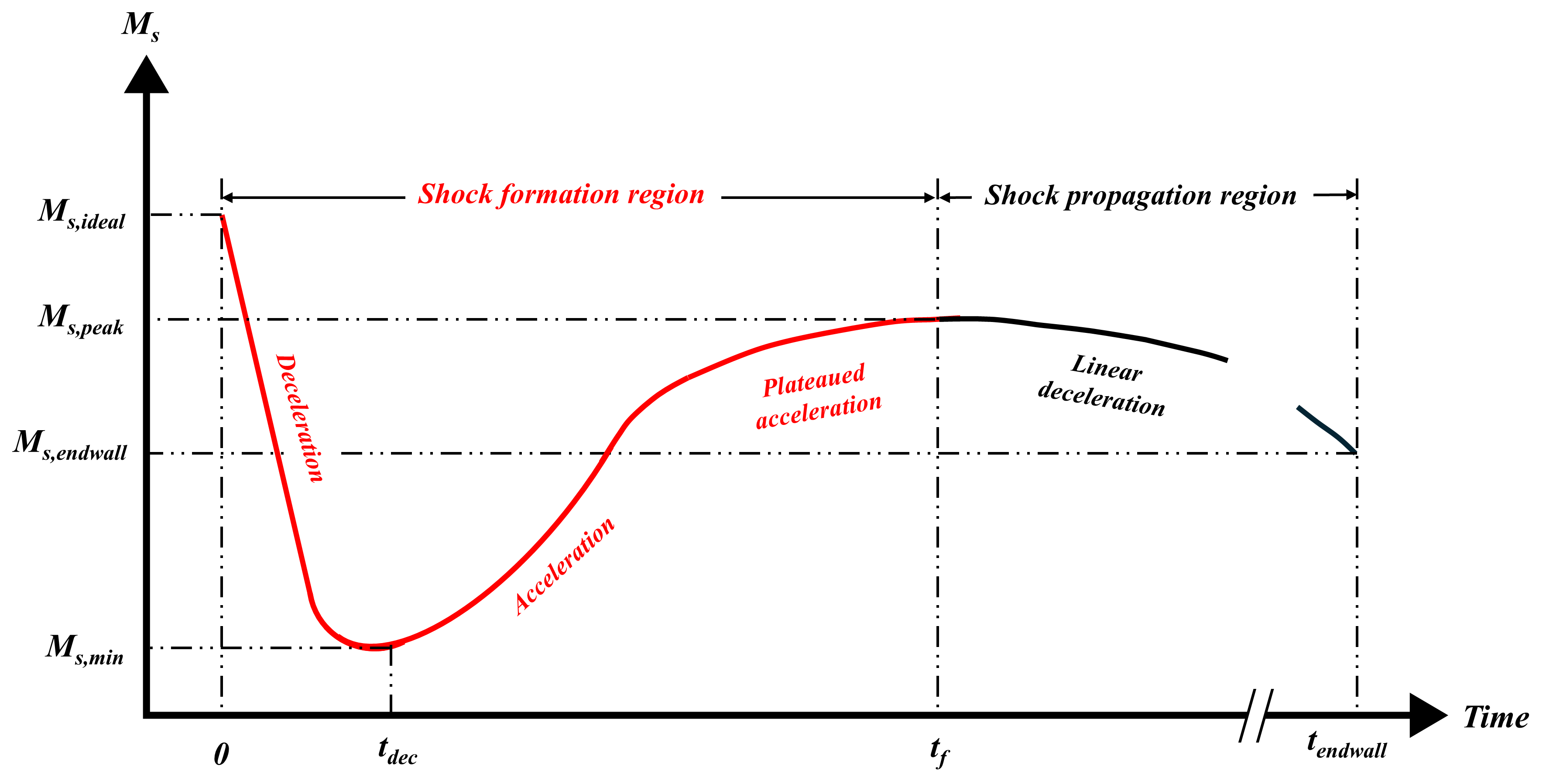}
    \caption{Schematic representation of the typical shock Mach number variation in the driven section of a single-diaphragm shock tube.} 
    \label{fig:Schematic_shock_formation}
\end{figure*}

\subsection{Representative shock trajectory in a single-diaphragm shock tube}
Based on the understanding of the shock formation process through combined experimental and numerical efforts, a representative profile of the shock Mach number is developed and shown in Fig. \ref{fig:Schematic_shock_formation} as a function of time. The profile exhibits an initial deceleration phase until the shock velocity reaches a minimum Mach number, $M_{s,min}$. This phase begins with the ideal shock velocity, $M_{s,ideal}$, which corresponds to the shock formation at the smallest diaphragm rupture. Here, the rupture acts as a miniature shock tube, creating a discontinuity at the orifice. The deceleration is attributed to radial expansion of the shock wave -- a rapid process lasting only a few hundred microseconds. During this phase, the diaphragm opening remains insufficient to counteract the deceleration or initiate acceleration. Gradually, the deceleration slows and transitions into an inflection point, referred to as $t_{dec}$ , where the shock begins to accelerate. This acceleration is driven by the increasing diaphragm aperture, allowing a larger flow of high-pressure driver gas, consistent with White's\cite{white1958influence} hypothesis that compression waves coalesce to accelerate the shock wave. The measured values of $t_{dec}$ for cases S1–S7 are listed in Table \ref{tab:shock_formation_table}. Beyond $t_{dec}$, the shock undergoes sustained acceleration, transitioning into a gradual plateau as it reaches its peak velocity, $M_{s,peak}$. This evolution occurs over a spatial extent termed the shock formation distance, $x_f$ , which defines the distance from the diaphragm to the point where the shock attains $M_{s,peak}$. The corresponding temporal parameter, the shock formation time, $t_f$ , represents the duration from diaphragm rupture to the completion of shock acceleration. The acceleration phase extends beyond the complete diaphragm opening, as the coalescence of compression waves requires finite time to catch up with the shock front. Subsequently, the shock begins to attenuate due to boundary layer growth. 

\section{Correlations for shock parameters in shock formation region}
To comprehensively analyze the shock formation process, the following subsections present a detailed framework for computing critical parameters observed in the shock Mach number profiles. First, the time until deceleration $(t_{dec})$ and the corresponding minimum Mach number $(M_{s,min})$ are correlated with numerical data, emphasizing the role of gas properties and geometric factors. Next, a theoretical model for predicting the peak shock Mach number, $M_{s,peak}$, is introduced, accounting for the influence of diaphragm opening dynamics and mass flow restrictions. Subsequently, the shock formation distance $(x_f)$ and formation time $(t_f)$ are determined, highlighting their dependence on pressure ratios and diaphragm opening times. Lastly, a comprehensive correlation is proposed to replicate the temporal evolution of the shock Mach profile, $M_s(t)$, during the deceleration and acceleration phases. This correlation provides a robust tool for understanding and predicting shock dynamics in the shock formation region. Together, these subsections establish a cohesive framework for quantifying the parameters governing shock formation.

\subsection {Shock deceleration time $(t_{dec.})$ and minimum shock Mach number $(M_{s,min})$}
\begin{figure*}
    \centering
    \includegraphics[width=0.99\textwidth]{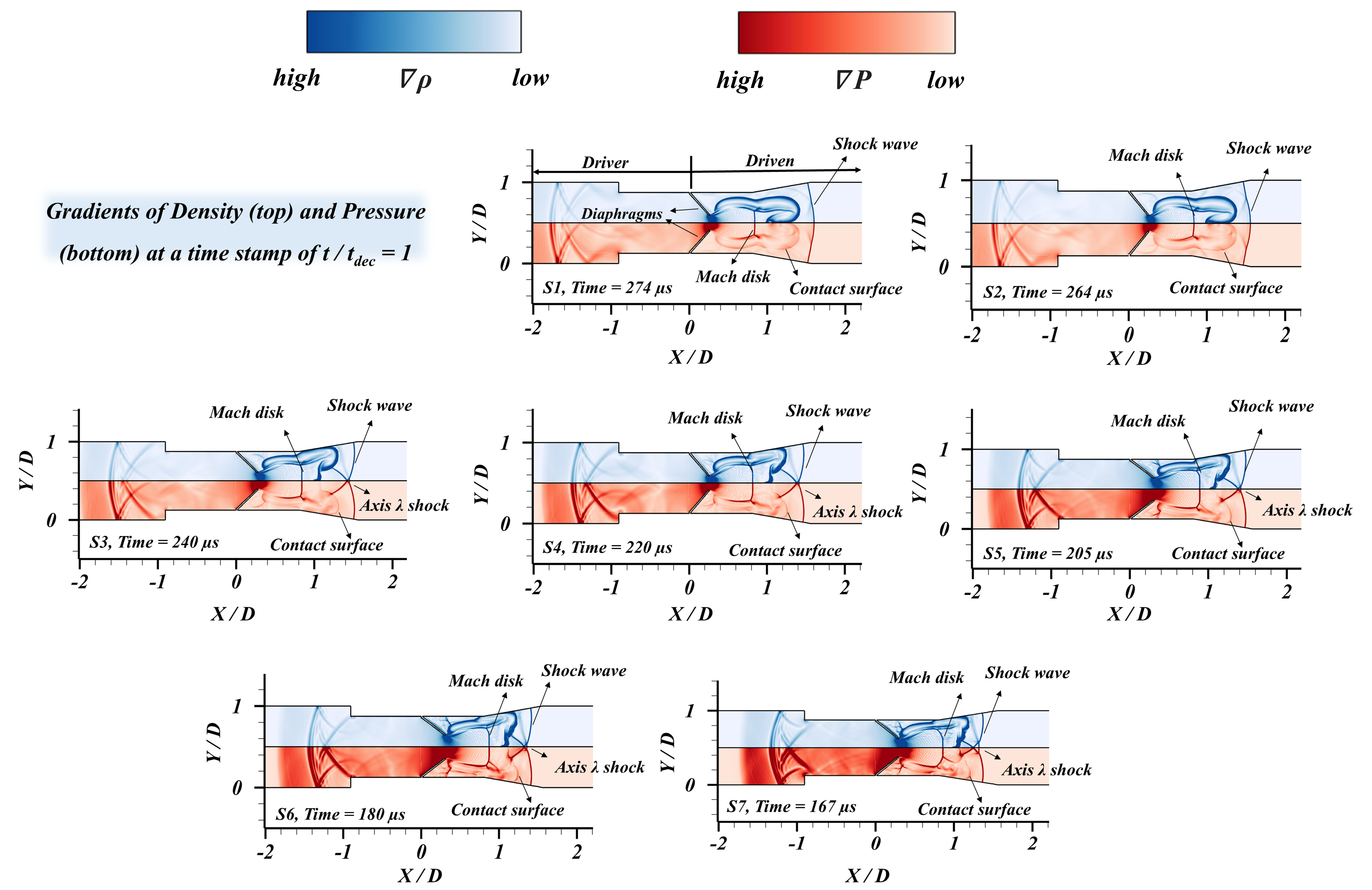}
    \caption{Gradients of density and pressure for cases S1 to S7 at different times since PD rupture, but at $\frac{t}{t_{dec}}$ = 1. Shock locations for all the cases at the end of shock deceleration is approximately the same} 
    \label{fig:Shock_formation_common_feature}
\end{figure*}

The time until the onset of shock deceleration, denoted as $(t_{dec.})$, represents a critical phase in shock dynamics immediately following the diaphragm rupture. During this phase, the shock wave undergoes rapid radial expansion, leading to a marked reduction in its strength. This behavior leads to a decrease in the shock Mach number to a minimum value, $(M_{s,min})$. Numerical simulations for all cases (S1–S7) revealed distinct trends in $(t_{dec})$ and $(M_{s,min})$, as summarized in Table \ref{tab:shock_formation_table}. Additionally, Fig. \ref{fig:Shock_formation_common_feature} illustrates the density (top half) and pressure (bottom half) gradients across the shock front at $\frac{t}{t_{dec}}$ = 1 time instant. A notable observation is that the shock positions at the end of the deceleration phase are nearly identical across all cases, suggesting that $(t_{dec})$ is not solely governed by the initial diaphragm rupture dynamics. Instead, it reflects a more complex interplay of driver-to-driven gas properties, such as the pressure ratio $P_{41}$ and $a_{41}$, and the geometric configuration of the driven section. This similarity in shock position highlights that while the intensity and duration of the deceleration phase vary among cases, the spatial scale over which the dynamics occur remains the same.

\begin{table*}
    \centering
    \begin{tabular}{cclcclllcc}
        \hline
         \# &  $P_{41}$&   $a_{41}$ &$t_{op,5 \rightarrow 100}$, $\mu$s
&$M_{s,ideal}$& $M_{s,peak}$& $M_{s,min}$&    $t_{dec.}$, $\mu$s&$t_f$, $\mu$s& $x_f/D$\\
         \hline  
         S1&  70&   2.43 &705
&3.22& 3.04& 1.75&       274&2180&   18.65\\
         S2&  75&   2.46 &655
&3.28& 3.08& 1.76&        264&2140&    18.50\\
         S3&  95&   2.57 &530
&3.51& 3.29& 1.78&        240&1980&    18.29\\
         S4&  118&   2.64 &430
&3.74& 3.51& 1.83&        220&1800&    17.81\\
         S5&  133&   2.69 &405
&3.86& 3.64& 1.96&        205&1660&    17.3\\
         S6&  151&   3.16 &360
&4.24& 4.00& 2.21&        180&1450&    17.22\\
         S7&  191&   3.16 &320&4.46& 4.21& 2.3&        167&1380&    16.92\\
         \hline
    \end{tabular}
    \caption{Measurements of peak and minimum shock Mach number ($M_{s,peak}$ and $M_{s,min}$), deceleration time of shock wave ($t_{dec.}$), shock formation distances ($x_f$) and shock formation times ($t_f$) from numerical simulations for cases S1 to S7}
    \label{tab:shock_formation_table}
\end{table*}

To quantify $(t_{dec.})$, a correlation dependent on $P_{41}$ and $a_{41}$ was developed, as expressed in Eq. \ref{eq:shock_deceleration_time}. This correlation is intentionally decoupled from the diaphragm opening time, as $(t_{dec})$ consistently exhibits values smaller than the diaphragm opening time (see Table \ref{tab:shock_formation_table}).

\begin{figure*}
    \centering
    \includegraphics[width=0.85\textwidth]{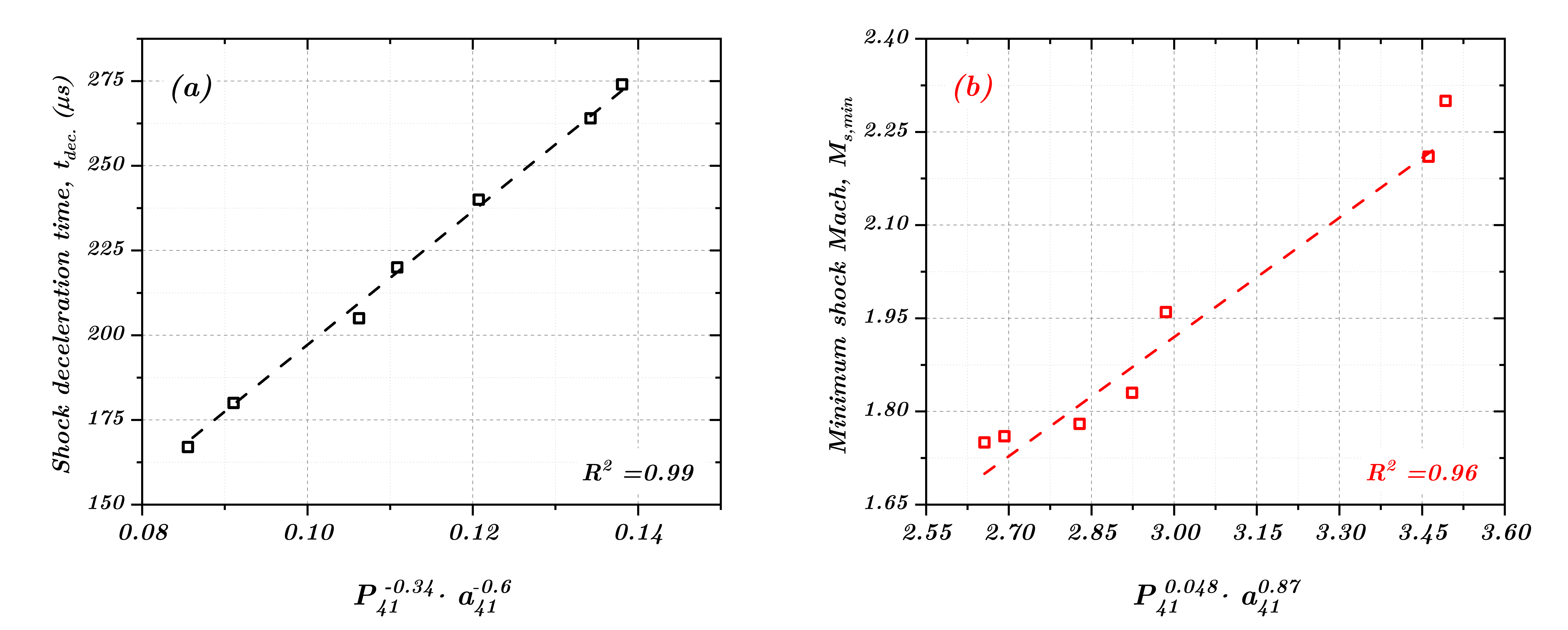}
    \caption{Data fits for (a) time of shock deceleration $(t_{dec.})$ and minimum shock Mach number $(M_{s,min})$.} 
    \label{fig:t_dec_and_Ms_min}
\end{figure*}

\begin{equation}
    t_{dec.}  = f(P_{41},a_{41}) =  d_1 \cdot P_{41}^{d_2} \cdot a_{41}^{d_3}
    \label{eq:shock_deceleration_time}
\end{equation}

where $d_1$,$d_2$,$d_3$ are fitted coefficients with values 1972, -0.34, and -0.6 respectively. Figure \ref{fig:t_dec_and_Ms_min}a shows that the proposed correlation demonstrates excellent agreement with simulation data, offering a predictive capability for determining $t_{dec}$ in similar shock tube configurations. Similarly, the $M_{s,min}$ can also be expressed as a correlated quantity depending on $P_{41}$ and $a_{41}$ shown as Eq. \ref{eq:shock_minimum}. This correlation also aligns well with the trends observed in numerical data, as illustrated in Fig. \ref{fig:t_dec_and_Ms_min}b. The values of fit coefficient for Eq. \ref{eq:shock_minimum} are $e_1$ = 0.64, $e_2$ = 0.048, $e_3$ = 0.87. These correlations provide a comprehensive framework for evaluating both $t_{dec}$ and $M_{s,min}$, offering valuable insights into the early stages of shock formation. 

\begin{equation}
    M_{s,min}  = f(P_{41},a_{41}) =  e_1 \cdot P_{41}^{e_2} \cdot a_{41}^{e_3}
    \label{eq:shock_minimum}
\end{equation}

\subsection{Peak shock Mach number ($M_{s,peak}$)}
It is well established that the gradual opening of the diaphragm limits the mass flow at the diaphragm location resulting in a reduction of shock strength. This constraint delays both the formation and acceleration of the shock wave, ultimately resulting in a lower $M_{s,peak}$. The effect is particularly pronounced in shock tube experiments operating under conditions where the driver-to-driven pressure ratio ($P_{41}$) is moderately high (approximately 500) and the driven pressure ($P_1$) is relatively elevated, on the order of 1 bar. In contrast, experiments conducted by White\cite{white1958influence} at hypersonic conditions (at Mach $\geq$ 5 and high $P_{41}$ -- achieved by significantly lowering ($P_1$)) showed a much higher $M_{s,peak}$ exceeding the predictions by ideal shock relations (Eq. \ref{eq:ideal_relation}) as well. Few more literature studies have also confirmed the same\cite{rothkopf1976shock,ikui1969investigations1}.

\begin{figure}
    \centering
    \includegraphics[width=0.47\textwidth]{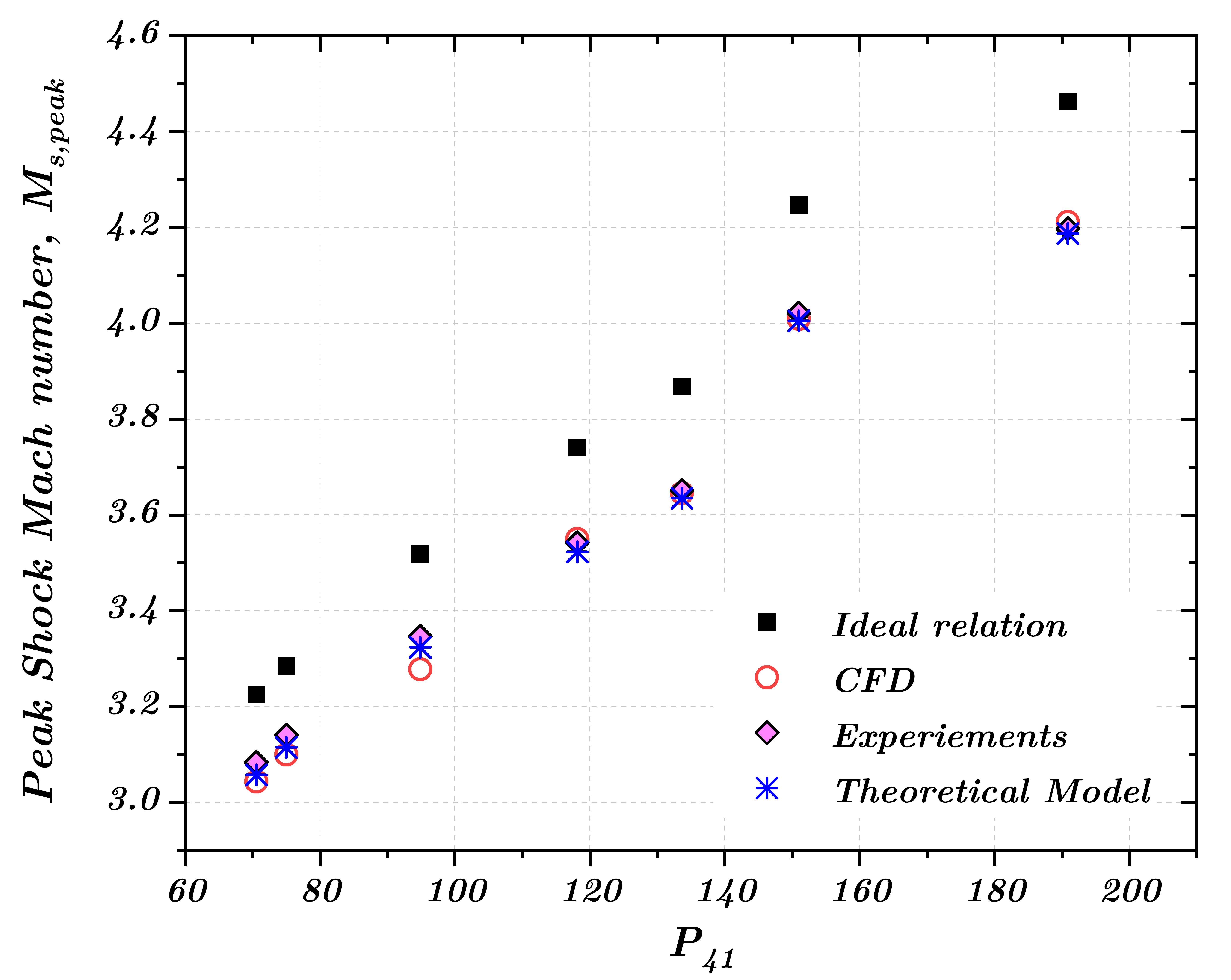}
    \caption{Comparison of peak shock Mach number $(M_{s,peak})$ from CFD, experiments and theoretical model} 
    \label{fig:P41_model_exp}
\end{figure}

In the present study, the experimentally measured and numerically simulated  $M_{s,peak}$ values are compared with the ideal shock velocity predicted by Eq. \ref{eq:ideal_relation}, as illustrated in Fig. \ref{fig:P41_model_exp}. The results demonstrate that the peak velocities are substantially lower than the ideal predictions (indicated by the solid black symbols). This discrepancy confirms that the observed behavior aligns with the scenario where gradual diaphragm opening restricts the mass flow, leading to a reduction in the peak shock velocities. 


A few theoretical models have been developed in the literature to predict $M_{s,peak}$. The models proposed by White\cite{white1958influence} and Ikui\cite{ikui1969investigations1} utilize a multi-stage framework that approximates unsteady isothermal compression during the early stages of shock formation. These approaches are particularly effective in scenarios with exceptionally high $P_{41}$, where the peak shock velocity exceeds the ideal prediction given by Eq. \ref{eq:ideal_relation}. At lower $P_{41}$, Pakdaman et al.\cite{pakdaman2016diaphragm}. proposed a shock formation model that integrates a diaphragm opening model with an empirical correlation derived from the work of Gaetani et al.\cite{gaetani2008shock}. While this model achieves good agreement with the experimental data, it remains fundamentally a curve fit to the empirical results. Alves et al.\cite{alves2021modeling} later extended this model to a simple fixed area orifice-type opening, similar to a constant slit-type configuration. Their approach involved equating the mass flow rate through the diaphragm orifice with the mass flow through the shock wave. To account for expansion effects and losses near the orifice, a discharge coefficient from literature was incorporated in mass flow computations.

In the current study, a direct extension of Alves et al.'s\cite{alves2021modeling} model may not be entirely appropriate due to the continuously increasing flow area as the diaphragm opens. Therefore, a new theoretical model has been developed, building upon the frameworks established by Pakdaman et al.\cite{pakdaman2016diaphragm} and Alves et al.\cite{alves2021modeling}. To account for the absence of a discharge coefficient for gradual diaphragm opening, a correction factor $(\frac{D_{mid}}{D_{Driven}})$ is introduced to adjust the mass flow calculation through the opened diaphragm, which fortunately provides a good fit to the data. Where $D_{mid}$ is the maximum flow width available after the diaphragm has opened in the square to round transition section between driver and driven (see schematic Fig. \ref{fig:Schematic_shock_tube}).  

It should also be noted that several key assumptions from the models proposed by Pakdaman et al.\cite{pakdaman2016diaphragm} and Alves et al.\cite{alves2021modeling} are retained in the present model and skipped for brevity. The sequence of computations in the theoretical model is outlined below.

The computation begins by iteratively estimating the Mach number of the expanding driver gas $M_3$ as it escapes through the increasing diaphragm aperture orifice as shown in \ref{eq:M_3}.
\begin{equation}
\frac{A_{max}}{A_{t}} = \frac{1}{M_3} \left( \frac{2}{\gamma_4 + 1} \left( 1 + \frac{\gamma_4 - 1}{2} M_{3}^2 \right) \right)^{\frac{\gamma_4 + 1}{2(\gamma_4 - 1)}}
\label{eq:M_3}
\end{equation}

Here, the term $\frac{A_{max}}{A_t}$, represents the inverse of fractional diaphragm opened area and can be approximated from the correlation Eq. \ref{eq:sigmoid}. $\gamma$ represents the ratio of specific heats, with subscripts used to differentiate the different regions in the shock tube. Region 4 refers to the unperturbed driver gas, while region 3 refers to the expanded driver gas. Once $M_3$ is determined for different opened areas, all flow variables in region 3 can be computed using isentropic flow relations, as shown in Eqs. \ref{eq:a_3}, \ref{eq:rho_3}, \ref{eq:P_3}, and \ref{eq:T_3} which give the speed of sound, pressure, density, and temperature, respectively in region 3:

\begin{equation}
a_3 = \frac{2 a_4}{(\gamma_4 - 1) M_3 + 2}
\label{eq:a_3}
\end{equation}

\begin{equation}
\rho_3 = \rho_4 \left( \frac{T_4}{T_3} \right)^{-\frac{\gamma_4}{\gamma_4 - 1}}
\label{eq:rho_3}
\end{equation}

\begin{equation}
P_3 = P_4 \left( \frac{a_3}{a_4} \right)^{\frac{2 \gamma_4}{\gamma_4 - 1}}
\label{eq:P_3}
\end{equation}

\begin{equation}
T_3 = T_4 \left( \frac{a_3}{a_4} \right)^{2}
\label{eq:T_3}
\end{equation}

Likewise the stagnation properties are computed from equations \ref{eq:P_03}, \ref{eq:T_03}, \ref{eq:rho_03}.

\begin{equation}
P_{03} = P_3 \left( 1 + \frac{\gamma_4 - 1}{2} M_3^2 \right)^{\frac{\gamma_4}{\gamma_4 - 1}}
\label{eq:P_03}
\end{equation}

\begin{equation}
T_{03} = T_3 \left( 1 + \frac{\gamma_4 - 1}{2} M_3^2 \right)
\label{eq:T_03}
\end{equation}

\begin{equation}
\rho_{03} = \rho_{04} \left( \frac{T_{04}}{T_{03}} \right)^{-\frac{\gamma_4}{\gamma_4 - 1}}
\label{eq:rho_03}
\end{equation}

With the driver region properties established, the next step involves calculating the flow properties at the diaphragm opening. This is done using the condition that the stagnation pressure in region 3 and at the diaphragm opening (denoted by subscript $j$) must be equal. The Eqs. \ref{eq:Pj}, \ref{eq:rhoj}, \ref{eq:vj} are used to estimate the pressure, density, and velocity at the diaphragm opening.

\begin{equation}
P_j = \frac{P_{03}}{\left(1 + \frac{\gamma_4 - 1}{2}\right)^{\frac{\gamma_4}{\gamma_4 - 1}}}
\label{eq:Pj}
\end{equation}

\begin{equation}
\rho_j = \frac{\rho_{03}}{\left( 0.5 + 0.5 \gamma_4 \right)^{-\frac{1}{\gamma_4 - 1}}}
\label{eq:rhoj}
\end{equation}

\begin{equation}
v_j = \sqrt{\gamma_4 \left(\frac{R_{uni} \times 10^3}{M_{\text{wt,4}}}\right) T_j}
\label{eq:vj}
\end{equation}

Finally, the mass flow rate $(m_j)$ through the diaphragm opening is computed using Eq \ref{eq:mj}. In this expression, $D_{mid}$ represents the maximum available flow width from driver to driven, which is determined to be 73 mm. This value is obtained by subtracting the twice of diaphragm thickness (1.6 mm) from the square-section width (76.2 mm) connecting driver and driven (see schematic Fig. \ref{fig:Schematic_shock_tube}). The depth of the shock tube is assumed to be 1 m, following the methodology outlined from Alves et al.\cite{alves2021modeling} to maintain consistency in the mass flow computation. To account for temporal area change of the diaphragm orifice, the term $\frac{A_t}{A}$ is included in Eq. \ref{eq:mj}.

\begin{equation}
\dot{m}_j = \rho_j v_j \left( \frac{A_t}{A} \times D_{mid} \right)
\label{eq:mj}
\end{equation}

Once $\dot{m}_j$ is computed, the flow properties in the shocked gas can be obtained using ideal shock relations by iteratively matching the flow rate just behind the shock wave $(\dot{m}_2)$ to $\dot{m}_j$. This procedure is similar to that outlined by Alves et al.\cite{alves2021modeling}. In the computation of $\dot{m}_2$, the flow area is assumed to be equal to the diameter of the driven section (0.1016 m) times a unit width, as shown in Eq. \ref{eq:m2}. $\rho_2$ and $u_2$ are the density and velocity of the \textit{shocked} gas which are obtained from ideal shock relations and by iteratively guessing the shock Mach number, until a match with $\dot{m}_j$ is achieved. As $\dot{m}_j$ changes with further opening of diaphragm, a profile of shock Mach number with $\dot{m}_j$ is reached, the peak of which occurs when the diaphragm is fully open (i.e., $\dot{m}_j$ is maximum) and this is referred to as $(M_{s,peak})$

\begin{equation}
\dot{m}_2 = \rho_{2} \cdot u_2 \cdot D_{Driven}
\label{eq:m2}
\end{equation}

The predictions of $(M_{s,peak})$ from the theoretical model, CFD simulations, and experimental measurements are shown in Fig. \ref{fig:P41_model_exp}. The theoretical model captures the peak shock velocity with reasonable accuracy, aligning closely with the experimental data across a range of $P_{41}$. While slight deviations exist, particularly at lower $P_{41}$, the model generally performs well.

\begin{figure*}
    \centering
    \includegraphics[width=0.9\textwidth]{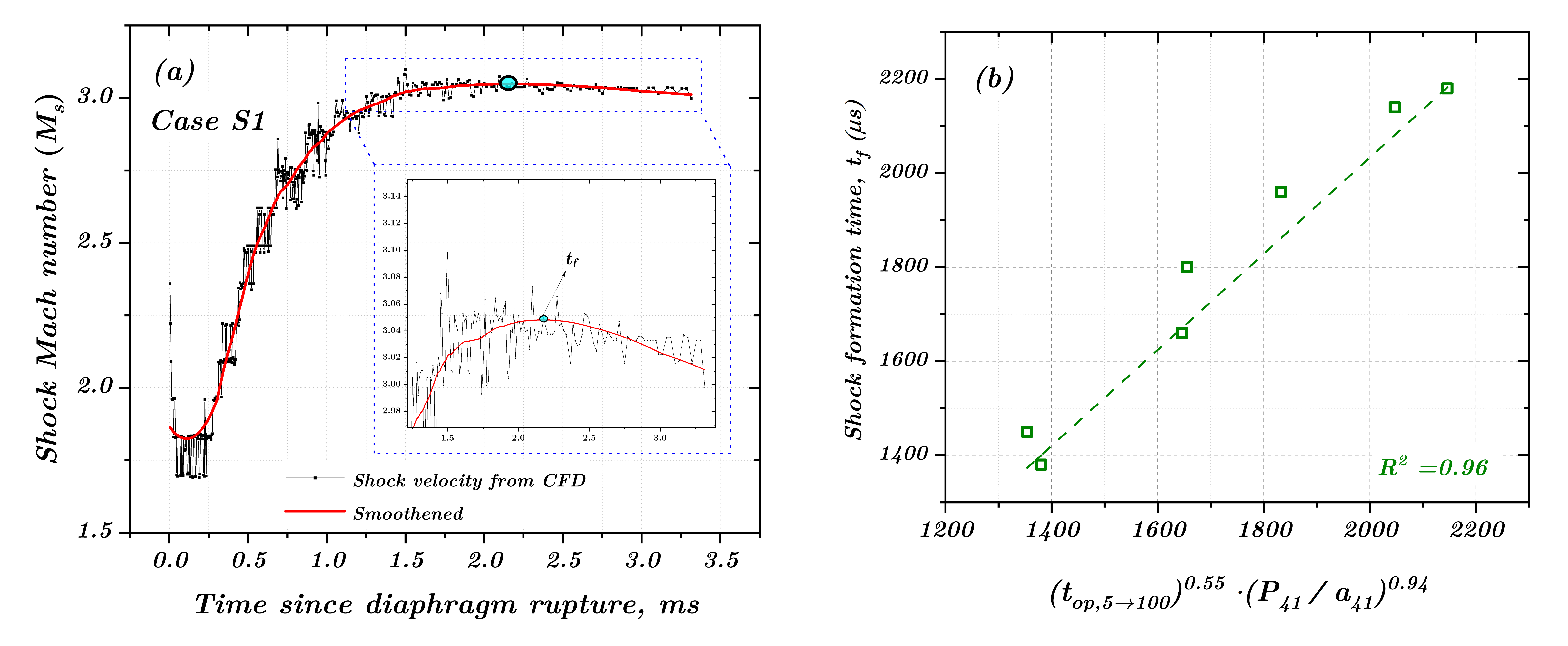}
    \caption{(a) Instantaneous shock velocity measurements for Case S1, highlighting the point where shock velocity begins to decrease, with a smoothed fit for clarity. The inset shows a detailed view around the deceleration onset. (b) Shock formation time $t_f$ from simulations, plotted against the correlation.} 
    \label{fig:Shock_formation_time}
\end{figure*}

\subsection {Shock formation distance and time}
The shock formation distance, ($x_f$), and shock formation time, ($t_f$), are critical parameters that define the spatial and temporal scales over which the shock reaches its fully developed state. $x_f$ is defined as the axial distance from the diaphragm to the point where the shock achieves its peak velocity, $M_{s,peak}$, while $t_f$ refers to the duration from diaphragm rupture to the onset of this peak velocity. These parameters are integral to understanding the performance and design constraints of shock tube experiments. Earliest measurements of $x_f$ were by White\cite{white1958influence} demonstrating a positive dependence of $x_f$ on $P_{41}$ or $M_s$. Later works on measuring of $x_f$ are by Rothkopf\cite{rothkopf1976shock} and Ikui\cite{ikui1969investigations1} and the conclusions essentially remained same from all the studies, and a dependence of $x_f$ on $P_{41}$ and $t_{op,100}$ (Eq. \ref{eq:shock_formation_distance}) is presented in literature. Interestingly, Rothkopf\cite{rothkopf1976shock} incorporated a dependence of the shock velocity on $x_f$, rather than directly on $P_{41}$. Since $P_{41}$ is a controlled input parameter, and the shock speed can be derived from it, using $P_{41}$ as a direct predictor of $x_f$ has been preferred in this study, as in the dimensional analysis of Ikui\cite{ikui1969investigations1}. Similar to $x_f$, the shock formation time ($t_f$) can also be computed, reflecting the time, since the diaphragm rupture, it takes for the shock to fully form. The relationship between $t_f$, $P_{41}$, and the diaphragm opening time $t_{op,100}$ can be expressed as in Eq. \ref{eq:shock_formation_time}.

\begin{equation}
    x_f  = K \cdot t_{op,100} \cdot P_{41}
    \label{eq:shock_formation_distance}
\end{equation}

\begin{equation}
    t_f  = t_{op,100} \cdot f(P_{41})
    \label{eq:shock_formation_time}
\end{equation}

Interpreting $x_f$ from Eq. \ref{eq:shock_formation_distance} requires careful consideration, as it suggests a direct dependence of $x_f$ on both $P_{41}$ and $t_{op,100}$. In White's\cite{white1958influence} experiments, shock formation distances increased over several meters as $P_{41}$ was exponentially increased ($10^2$ - $10^6$), which was achieved by lowering $P_1$, while $P_4$ remained largely unchanged. As a result, the diaphragm opening time $(t_{op,100})$ did not change significantly, meaning that the increase in $x_f$ could be directly attributed to the rise in $P_{41}$ (due to the reduction in $P_1$). A similar observation was made by Kashif et al.\cite{kashif2024insights} who followed White's\cite{white1958influence} methodology of increasing $P_{41}$ (13 - 80) by lowering $P_1$ and observed that $x_f$ varied by almost a meter (see Fig. 3a from Kashif et al.\cite{kashif2024insights}).

In contrast, in experiments where higher $P_1$ is also desirable, Simpson et al.\cite{simpson1967effect} highlighted that increasing $P_4$ while maintaining a constant $P_1$ results in less variation in shock formation distances. This is because an increase in $P_4$ decreases the diaphragm opening time $t_{op,100}$. Thus, there is a competing effect between $t_{op,100}$ and $P_{41}$ in determining $x_f$, as expressed in Eq. \ref{eq:shock_formation_distance}. This competition is also illustrated in Fig. 3b from Kashif et al.\cite{kashif2024insights}, where a variation of $P_4$ solely ($P_1$ = 0.13 bar), results in an almost constant $x_f$. Simpson et al.\cite{simpson1967effect} performed experiments at same $P_{41}$ but used diaphragms of two different materials, resulting in contrasting opening times. As a result the authors observed varying values of $x_f$ (in Fig. 5 of Simpson et al.\cite{simpson1967effect}) Notably, the case with the faster diaphragm opening produced a smaller $x_f$. They also calculated the shock formation time using the relationship $t_f = \frac{x_f}{\text{max. speed}}$.

In the present study, as shown in Table \ref{tab:Experimental conditions}, $P_1$ is held constant while $P_4$ is increased, leading to a reduction in $t_{op,100}$. Consequently, significant variations in $x_f$ were not observed. To confirm this, both $x_f$ and $t_f$ were computed from numerical simulations and are tabulated in Table \ref{tab:shock_formation_table}. Figure \ref{fig:Shock_formation_time}a illustrates the computation of $t_f$. The instantaneous shock velocity has oscillations (black trace) and hence a smoothening operation with a moving mean of window size 3 (red trace) was performed to discern the acceleration with deceleration. The time instant when the shock velocity begins to decrease is labeled as $t_f$ in Fig. \ref{fig:Shock_formation_time}a. The corresponding shock formation distance, $x_f$, is identified by locating the shock at the time $t_f$. As expected, the variation in $x_f$ across cases is minimal but follows a trend. However, $t_f$ decreases by nearly 50 \% moving from S1 to S7. A correlation (Eq. \ref{eq:shock_formation_time_corr}) was fitted to the measured values of $t_f$, as shown in Fig. \ref{fig:Shock_formation_time}b. $t_f$ is considered to be a function of $t_{op,5 \rightarrow 100}$ and not $t_{op,100}$ as the diaphragm opening time in simulations was chosen to be $t_{op,5 \rightarrow 100}$, as described in the methodology section.

\begin{equation}
    t_f  = f(P_{41},a_{41},t_{op,5 \rightarrow 100}) = \left ({\frac{P_{41}}{a_{41}}} \right)^{f_1} \cdot t_{op,5 \rightarrow 100}^{f_2}
        \label{eq:shock_formation_time_corr}
\end{equation}

Since variation in $x_f$ is minimal no correlation was developed, in the subsequent section, a comparison is drawn between numerical shock formation distance and the one obtained with a correlated velocity profile is provided.

\begin{figure*}
    \centering
    \includegraphics[width=0.85\textwidth]{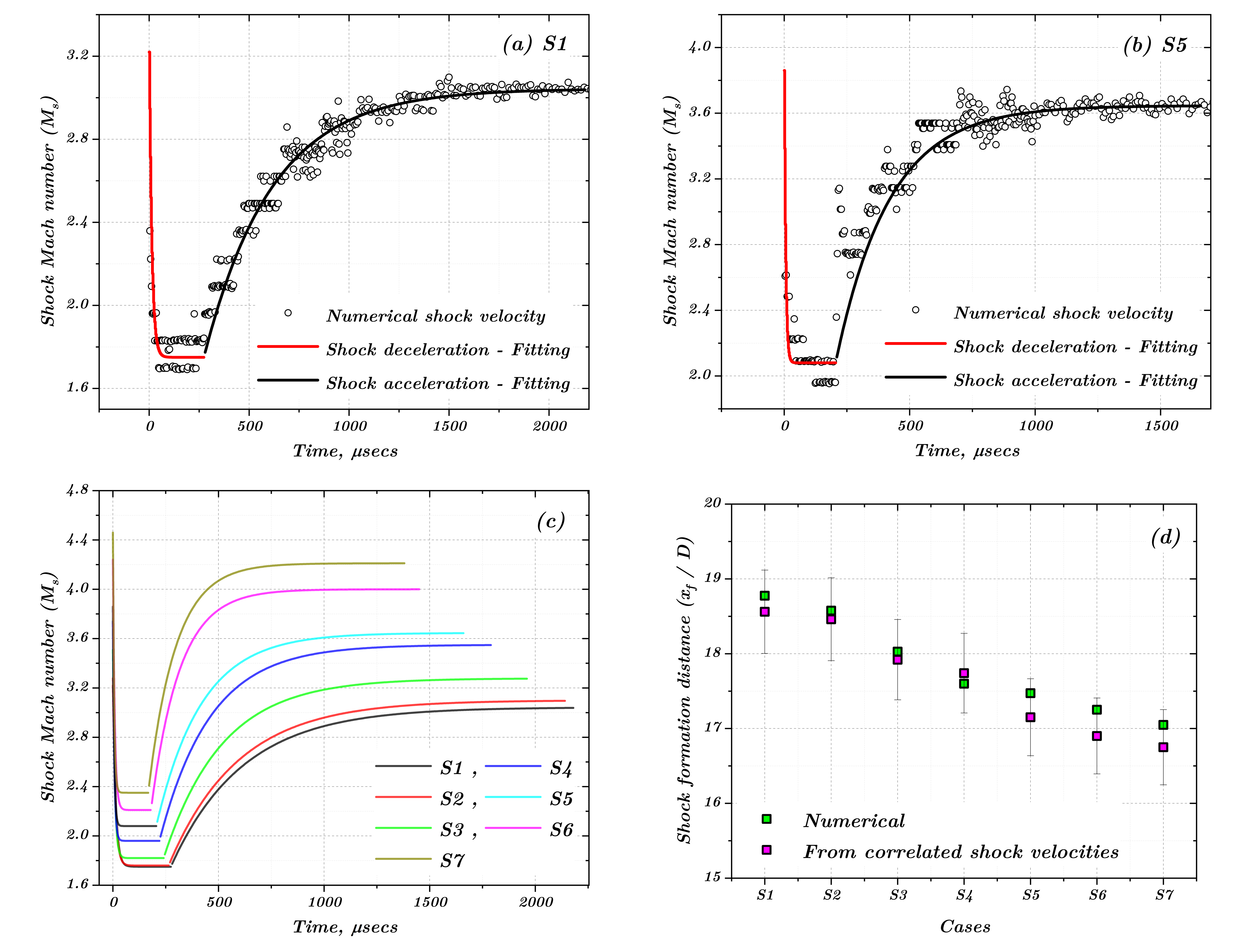}
    \caption{Comparisons of numerically measured shock velocities and predictions of correlations for (a) S1 and (b) S5, (c) A plot showing the time-Mach graph of the shockwave obtained from correlations, (d) Comparison of shock formation distances $x_f / D$ obtained from numerical data and correlated shock velocity profiles across all cases, highlighting the agreement between both methods.} 
    \label{fig:Shock_profile_fitting}
\end{figure*}

\subsection{Shock Mach number profile}
The evolution of the shock velocity during its formation phase reflects a complex interplay of radial expansion and compression wave interactions. Capturing this temporal variation accurately is essential for characterizing shock formation dynamics and optimizing shock tube design. In this study, a generalized correlation is developed to predict the shock Mach profile \(M_s(t)\) during the formation process. This correlation is developed based on the observed trends in deceleration and acceleration phases, as illustrated schematically in Fig. \ref{fig:Schematic_shock_formation}, and is validated against numerical simulations. The shock velocity profile is modeled as a piecewise function to account for the distinct physical mechanisms dominating different phases of the formation process:

\begin{enumerate}
    \item \textbf{Deceleration Phase (\(t < t_{dec}\)):} During this phase, the shock velocity decreases from \(M_{s,\text{ideal}}\) to \(M_{s,\text{min}}\), driven by the rapid radial expansion of the shock wave near the diaphragm. The correlation for this phase is expressed as (Eq. \ref{eq:fitted_velocity_dec}):
    \begin{equation} \small
        M_s(t) = M_{s,\text{min}} + (M_{s,\text{ideal}} - M_{s,\text{min}}) \cdot \exp\left(- (P_{41}^{\alpha}) \cdot (a_{41}^{\beta}) \cdot \frac{t}{t_{dec}}\right)
        \label{eq:fitted_velocity_dec}
    \end{equation}
    
    \item \textbf{Acceleration Phase (\(t_{dec} \leq t \leq t_f\)):} Following the deceleration phase, the shock velocity increases as compression waves coalesce and the diaphragm opens further, enabling greater mass flow into the driven section. The correlation for this phase is given by (Eq. \ref{eq:fitted_velocity_acc}):
    \begin{equation} \small
        M_s(t) = M_{s,\text{min}} + (M_{s,\text{peak}} - M_{s,\text{min}}) \cdot \left(1 - \exp\left(-(P_{41}^{\gamma}) \cdot (a_{41}^{\delta}) \cdot \frac{t - t_{dec}}{t_f - t_{dec}}\right)\right)
        \label{eq:fitted_velocity_acc}
    \end{equation}
\end{enumerate}

In these expressions, \(\alpha\), \(\beta\), \(\gamma\), and \(\delta\) are empirically determined parameters that encapsulate the effects of gas properties (\(P_{41}\) and \(a_{41}\)) and diaphragm dynamics on the shock formation process.

The shock velocity profiles predicted by the proposed correlation were compared to the results obtained from numerical simulations for all cases (S1 – S7). Representative comparisons for cases S1 and S5 are presented in Figs. \ref{fig:Shock_profile_fitting}(a) and (b), respectively. Fig. \ref{fig:Shock_profile_fitting}(c) further extends the comparison to include all cases. The strong agreement between the predicted and simulated shock velocity profiles highlights the effectiveness of the proposed correlation in accurately capturing the temporal evolution of shock velocity. Furthermore, the shock formation distances predicted using the correlated velocity profiles were compared to those derived from numerical simulations. As illustrated in Fig. \ref{fig:Shock_profile_fitting}(d), the predicted shock formation distances exhibit a close match with the simulation data, demonstrating the capability of the correlation to represent both the temporal and spatial characteristics of shock formation dynamics.

\section{Conclusions and future work}
The present study provides a comprehensive understanding of shock wave formation and propagation dynamics in single-diaphragm shock tubes by combining experimental observations with detailed computational simulations. High-speed imaging and diaphragm opening measurements underscore the significant role of diaphragm mechanics in influencing shock wave intensity and formation. Numerical simulations have offered deeper insights into the complex flow phenomena following diaphragm rupture, validating the critical impact of initial boundary conditions on shock dynamics. The key findings are summarized below:
\begin{itemize}
    \item The diaphragm opening time comprises of two phases: The stretching and tearing phase, denoted by $t_{op,5}$, and the pure bending phase, denoted by $t_{op,5 \rightarrow 100}$. $t_{op,5}$ linearly varies with the diaphragm thickness and driver pressure while $t_{op,5 \rightarrow 100}$ is primarily a function of the driver pressure. A sigmoidal model accurately captures the aperture's opening evolution.
    
    \item The wave system immediately after diaphragm burst comprises of complex shock and expansion wave structures, including Mach disks and $\lambda$-shocks. Shock planarity is achieved sooner at higher pressure ratios, with oscillations reducing further downstream as the shock stabilizes.
    
    \item The shock exhibits an initial deceleration phase, followed by acceleration to a peak Mach number, and transitions to attenuation due to boundary layer effects. New parameters, such as shock deceleration time ($t_{dec}$) and minimum shock Mach number ($M_{s,min}$), are introduced to characterize the early shock dynamics.

    \item Peak shock Mach number, $M_{s,peak}$ is consistently lower than ideal predictions due to diaphragm-induced mass flow limitations. Minimum shock Mach number, $M_{s,min}$ highlights significant deceleration during initial shock formation and is dependent on diaphragm pressure ratio and ratio of sound speeds.

    \item Correlations have been developed to predict the temporal evolution of the shock Mach number profile, incorporating diaphragm opening dynamics, pressure ratios, and mass flow effects. The correlations effectively replicate numerical trends, providing predictive capabilities for varying shock tube configurations.
\end{itemize}

The proposed correlations provide a practical framework for predicting the shock Mach number profile across a range of experimental conditions. By accounting for the underlying physical mechanisms in both the deceleration and acceleration phases, the correlation offers a unified approach for characterizing shock formation dynamics. This framework serves as a valuable tool for the shock tube community, enabling accurate predictions of shock behavior and informed design of shock tube experiments. Looking forward, the refinement of these models to incorporate more complex geometries and operational scenarios will extend their applicability to diverse shock tube configurations. Future work will also focus on exploring the interplay between diaphragm mechanics and shock attenuation in turbulent boundary layers. By bridging experimental insights with theoretical advancements, this study contributes to the broader understanding of shock tube physics and establishes a foundation for optimizing shock wave experiments and facilities.

\begin{acknowledgments}
This work was sponsored by King Abdullah University of Science and Technology (KAUST) and supported by the KAUST Supercomputing Laboratory (KSL). All simulations were performed on KSL’s Shaheen III supercomputer. Convergent Science provided CONVERGE licenses and technical support for this work. The authors appreciate Md Zafar Ali Khan's assistance in the shock tube experiments.
\end{acknowledgments}

\section*{Data Availability Statement}
The data that support the findings of this study are available from the corresponding author upon reasonable request.

\bibliographystyle{unsrt}
\bibliography{aipsamp}

\end{document}